\documentclass[
aps
prd,
preprint,
superscriptaddress,
amsfonts,
amssymb,
amsmath,
preprintnumbers,
floatfix,
nofootinbib,
unsortedaddress,
showpacs,
tightenlines,
floats,
a4paper,
eqsecnum,
]{revtex4}
\def\dhc{\bar{\delta}_\m{hc}}
\def\Ki{K_\m{i}(r)}

\def\al{\alpha}
\def\be{\begin{equation}}
\def\ee{\end{equation}}
\def\ep{\epsilon}
\def\pa{\partial}

\def\fr{\frac}
\def\de{\delta}
\def\ga{\gamma}
\def\de{\delta}
\usepackage{comment}

\def\al{\alpha}
\usepackage[dvipdfmx]{graphicx}
\def\be{\begin{equation}}
\def\ee{\end{equation}}
\def\ep{\epsilon}
\def\pa{\partial}

\def\fr{\frac}
\def\de{\delta}

\def\ga{\gamma}
\def\Ga{\Gamma}
\def\de{\delta}

\def\al{\alpha}
\def\si{\sigma}

\def\m{\mathrm}

\def\ri{r_\m{i}}

\makeindex

\begin{document}

\title{Identifying the most crucial parameters of the initial curvature profile for
primordial black hole formation
%Precise Determination of Primordial Black Hole Formation Condition for 
%Various Initial Curvature Profiles
}
\author{Tomohiro Nakama}

\affiliation{Department of Physics,
Graduate School of Science,\\ The University of Tokyo, Bunkyo-ku,
Tokyo 113-0033, Japan
}

\affiliation{Research Center for the Early Universe (RESCEU),\\
Graduate School of Science, The University of Tokyo, \\ Bunkyo-ku,
Tokyo 113-0033, Japan
}

\author{Tomohiro Harada}

\affiliation{Department of Physics, 
Rikkyo University, Toshima, Tokyo 175-8501, Japan
}

\author{A.~G.~Polnarev}

\affiliation{Astronomy Unit, School of Physics and Astronomy, \\
Queen Mary
University of London, \\ Mile End Road, London E1 4NS, United Kingdom
}

\author{Jun'ichi Yokoyama}

\affiliation{Research Center for the Early Universe (RESCEU),\\
Graduate School of Science, The University of Tokyo, \\ Bunkyo-ku,
Tokyo 113-0033, Japan
}

\date{\today}

\affiliation{Kavli Institute for the Physics and Mathematics 
of the Universe (Kavli IPMU), WPI, TODIAS,\\
The University of Tokyo, Kashiwa, Chiba 277-8568, Japan}

\date{\today}

\preprint{RESCEU-45/13, RUP-13-11}

\date{\today}

\small

\begin{abstract}
Primordial black holes (PBHs) are an important tool in cosmology to probe
 the primordial spectrum of small-scale curvature perturbations that reenter
the cosmological horizon during radiation domination epoch.  We numerically solve the
evolution of spherically symmetric highly perturbed configurations to clarify the
criteria of PBHs formation using an extremely wide class of curvature profiles characterized by
five parameters, (in contrast to only two parameters used in all previous papers) which specify 
the curvature profiles not only at the
central region
but also at the outer boundary of configurations.  It is shown that formation
or non-formation of PBHs is determined essentialy by only two master parameters one of which
can be presented as an integral of curvature over initial configurations 
and the other is presented 
in terms of the position of the boundary and the edge of the core.

\end{abstract}
%\section{A few modifications to the paper by \\May and White}
%$E$ is used instead of $\ep$, and e represents energy density.\\
%$w=1+E+P/\rho$ is modified to $w=E+P/\rho$, where $E\equiv e/\rho$.
%\section{章のまとめ}
\maketitle
%%%%%%%%%%%%%%%%%%%%%%%%%%%%%%%%%%%%%%%%%%%%%%%%%%%%%%%%%%
\section{introduction}
It is well known that a region with large amplitude curvature profile
 can collapse to a primordial black hole (PBH)
 \cite{Zel'dovich-1974,Hawking:1971ei}. 
The probability of existence of such regions and hence the abundance of PBHs depends
 on the physical conditions in the very early universe. PBHs are formed soon
 after the region enters the cosmological horizon during the
 radiation-dominated epoch. 
%It is also known that in a radiation-dominated universe the
 %contribution of PBHs into the density of the universe being extremely
 %small at the moment of PBHs formation drastically increases in the
 %course of subsequent cosmological expansion. 
Even if the contribution of the energy density of PBHs to
the total cosmic energy density is extremely small at formation,
it drastically increases in the course of subsequent cosmic expansion
during radiation domination.
For this reason PBHs can be used as a unique and powerful tool for the exploration of very small-scale structure of the early universe. 
This tool is really unique since there are no alternative techniques to study this small-scale structure. 

PBHs with different masses have different cosmological and astrophysical
implications.  
 The PBHs with mass smaller than $\sim 10^{15}$g are especially interesting  
 because by now they would have evaporated 
through Hawking radiation \cite{Hawking:1974rv} and their abundance
is constrained by the effects of emitted high-energy particles and photons on the 
Big-Bang Nucleosynthesis \cite{Zel'dovich:1977sta,Novikov:1979pol,
1978AZh....55..231V,1978PAZh....4..344V,Miyama:1978mp,Kohri:1999ex}, 
 the gamma-ray
background \cite{Page:1976wx,MacGibbon:1987my,MacGibbon:1991vc},
galactic and extragalactic antiprotons \cite{1976ApJ...206....8C}
as well as on the cosmic microwave background radiation (CMB).

PBHs with larger masses would still be present and are
constrained by dynamical and lensing effects \cite{Paczynski:1985jf} 
and by the stochastic gravitational wave background
\cite{Saito:2008jc,Saito:2009jt}.  
All these constraints are updated and summarized in
\cite{Carr:2009jm}. 
For the mass range between $6\times 10^{20}$g and $4\times 10^{24}$g 
the most stringent
constraint is given by the cosmic dark matter density \cite{Griest:2013aaa}.
%In other words, PBHs with this mass range can entirely explain the 
%cold dark matter.

Even if PBHs would never have been detected these constraints (being
reformulated in terms of the constraints on the amplitude of small-scale perturbations)
will provide valuable information on inflationary cosmological models
\cite{1989LNP...332..369P,Sato:1980yn,Guth:1980zm,Starobinsky-1980}, which predict
formation of super-horizon curvature perturbations
\cite{Mukhanov:1982nu,Guth:1982ec,Hawking:1982cz,Starobinsky:1982ee}.
So far  their large-scale components have been precisely tested by
observations of CMB \cite{Hinshaw:2012aka,Ade:2013uln} and
large-scale structure.  It is important to probe 
the perturbation spectrum on significantly smaller scales as well in order to obtain
more helpful information to single out the correct 
inflation model.  Indeed there exist a number of 
inflation models that predict features at some small scales
\cite{PhysRevD.54.6040,
PhysRevD.42.3329,PhysRevD.50.7173,yokoyama-1997-673,
PhysRevD.58.083510,Jun'ichi1998133,kawasaki-1999-59,
PTPS.136.338,1475-7516-2008-06-024,PhysRevD.59.103505,
PhysRevD.63.123503,
PhysRevD.64.021301,Kawasaki:2007zz,Kawaguchi:2007fz}
which may lead to PBHs abundances not compatible with observational constraints.
So we can gain
insights into primordial perturbations on small scales even from the absence of observed PBHs.
Or if PHBs are detected in the future it will extend our insight into inflationary models drastically.

In order to consistently use PBHs as a cosmological tool to probe the spectrum of small-scale curvature perturbations
, the relation between their mass function and the primordial perturbation spectrum
must be clarified. This enterprise involves several steps. First we must clarify the formation
condition of PBHs as a functional of the primordial curvature profiles of perturbed regions. Second
we must calculate the resultant mass of a PBH for each given profile. Finally we must calculate
the probability distribution functional of curvature perturbation spectrum which gives realization
probability of curvature profiles leading to PBH formation. This paper is predominantly devoted to the first issue and touches the second one in a preliminary manner. 
We will consider the second issue in more detail in another paper.

Originally the problem of PBH formation was studied analytically by considering the balance between gravity and
pressure gradient which hampers contraction, yielding a simple analytic criterion of PBH formation
 \cite{Carr:1974nx,Carr:1975qj}
\be
\fr{1}{3}\lesssim \dhc,\label{classical}
\ee
where $\dhc$ is the energy density perturbation averaged over the 
overdense region evaluated at the time of horizon crossing. 
This criterion has long been used in papers on
theoretical 
prediction of PBH abundance
(but has recently been refined in \cite{Harada:2013epa}).
In this simple picture, 
since $\dhc$ is the density perturbation averaged over the overdense region, 
the dependence on the 
profile or shape of perturbed regions has not been taken into account.

However, recent numerical analyses have shown that the condition for PBH formation does
 depend on the profile of perturbation
 \cite{Shibata:1999zs,Polnarev:2006aa} (see also \cite{1979STIN...8010983N,PhysRevD.59.124013}). 
Both \cite{Shibata:1999zs} and \cite{Polnarev:2006aa} used
two-parameter families of the initial profile (one corresponding to the amplitude of overdensity and
the other to the width of the transition region) and obtained two parametric conditions of PBH formation.  
It was clear from the above publications that one parametric description was not sufficient. However it was not clear whether the two-parametric description is good enough (say, in comparison with three-parametric description).
In the present paper, we extend these preceding analyses by making much more numerical
computations of PBH formation based on the initial curvature profile including many more parameters,
adopting the five-parameter family of profiles. We show that the criterion
 of PBH formation can still be expressed in terms of two
 crucial 
(master) parameters which represent the averaged amplitude of over
 density
 in the central region and the width of transition region at 
outer boundary, even though the considered profiles belong to
the five-parametric
 family.
Thereby we have obtained extra evidence that the two-parametric
 description mentioned above is self-consistent and 
provided a reliable physical interpretation of the 
criterion we have obtained.

\if
They introduced some functions to model primordial perturbed regions 
and obtained conditions for PBH formation for those perturbations
represented by their functions. 
Though there results were new and important, there seems to be some room 
for improvements. 
Firstly, physical interpretation of their condition of PBH formation is 
not clear and therefore 
what kinds of physical mechanisms play important roles in the process of 
PBH formation can not be understood. 
Secondly, they used functions which include at most two parameters and therefore
types of initial configuration shapes investigated are limited. 
Therefore their condition for PBH formation is applicable only to 
limited types of perturbation shapes. 
In reality, various kinds of perturbations must have been produced during inflation so 
in this paper we consider a wider class of shapes of perturbations 
by intoducing a function which includes five 
parameters, which can express a far more variety of perturbation shapes, 
thereby enabling more realistic analysis 
of PBH formation condition. for this extended class of shapes, we found 
a condition for PBH formation, which is expressed by two quantities 
characterizing profiles of perturbations and whose physical
interpretation 
can also be provided. 
The condition we obtained is a lot more general and accurate than 
those obtained before. 
\fi

In our previous paper \cite{1475-7516-2012-09-027} (hereafter PNY), 
we investigated the time evolution 
of spherical perturbed regions embedded in a flat 
Friedmann-Lemaitre-Robertson-Walker (FLRW) universe 
while the region is 
outside the cosmological horizon, 
using an asymptotic expansion. 
In order to calculate
the growth of the perturbed region 
after the horizon reentry, 
the Einstein equations need to be 
solved numerically.
In this paper, 
we first present the results of the 
numerical computation in the slicing 
used in our previous work, with the initial data
generated using the asymptotic expansion. %\textbf{イントロでぬるにも触れるべき}
In this slicing, at spatial infinity the time coordinate coincides with 
that in the background FLRW universe , 
so this slicing is sometimes referred to as cosmic time slicing 
in the literature.

%If the amplitude of the perturbation is small, 
%the pressure gradients dominates over gravity and
%the perturbation starts acoustic oscillation 
%after the perturbed region reenters the horizon.
%On the other hand, in case the amplitude is sufficiently large, 
%gravity dominates and the perturbed region collapses to 
%form a PBH.

During the numerical computation a singularity appears at the centre
after 
the apparent horizon is formed 
and as a result the computation has to be stopped and the subsequent
accretion 
onto the PBH cannot be followed. 
That is, the eventual mass of the PBH cannot be obtained. In order to
avoid 
this problem, 
we employ what is called the null slicing, which is also discussed in this paper. 
%and the physical interpretation of this condition is clear as will be shown. 

Finally, note that 
PBHs are formed only from extremely high peaks of perturbation,
corresponding 
to the tail of 
the probability distribution of primordial perturbation. 
The shape of these peaks has been calculated (see, for example,
\cite{Doroshkevich, Bardeen:1985tr}) 
and has turned out to be nearly spherically symmetric and monotonic near the peak. 
So in this paper we continue to consider only the spherically symmetric profiles which are monotonic near
the centre.

The rest of the paper is organized as follows. 
In \S I\hspace{-.1em}I, we briefly discuss the initial condition in terms of the asymptotic expansion
of the Einstein equations. In
 \S I\hspace{-.1em}I\hspace{-.1em}I
we present the basic equations used in the numerical computations.
Then in \S I\hspace{-.1em}V we review the time evolution of the perturbation in the cosmic time slicing 
.
In \S V we present the two parameters crucial for PBH formaiton. 
\S V\hspace{-.1em}I 
is devoted to
the determination of the PBH masses using the null slicing
and \S V\hspace{-.1em}I\hspace{-.1em}I to conclusion and discussion.

\section{Setting up the initial condition}
Assuming spherical symmetry, it is convenient to divide the collapsing matter
into a system of concentric spherical shells and to label each shell with a
Lagrangian comoving radial coordinate $r$. Then the metric can
be written in the form used by Misner and Sharp \cite{Misner:1964je}:
\begin{equation}
ds^2=-a^2dt^2+b^2dr^2+R^2(d\theta^2+\sin^2\theta d\phi^2),\label{1}
\end{equation}
where $R$, $a$ and $b$ are functions of 
$r$ and the time coordinate $t$. We consider a perfect fluid with the energy density $\rho(r,t)$ and pressure 
$P(r,t)$ 
and a constant equation-of-state parameter $\ga$, $P(r,t)=\ga \rho(r,t)$.
We express the proper time derivative of $R$ as
\begin{equation}
U\equiv \frac{\dot{R}}{a},\label{2}
\end{equation}
with a dot denoting a derivative with respect to $t$.%, we derive equations of motion for these variables as follows.

We define the mass, sometimes referred to as the Misner-Sharp mass in the literature, within the shell of circumferential radius $R$ by
\be
M(r,t)=4\pi\int^{R(r,t)}_0\rho(r,t)R^2dR.\label{defofM}
\ee
We consider the evolution of a perturbed region embedded in a 
flat Friedmann-Lemaitre-Robertson-Walker (FLRW) Universe with metric
\be
ds^2=-dt^2+S^2(t)(dr^2+r^2d\theta^2+r^2\sin^2\theta d\phi),
\ee
which is a particular case of (\ref{1}). The scale factor in this background evolves as
\begin{equation}
S(t)=\left(\fr{t}{t_\mathrm{i}}\right)^{\alpha} ,\quad\alpha\equiv\frac{2}{3(1+\gamma)},
\label{S0anddefofalpha}
\end{equation}
where $t_\mathrm{i}$ is some reference time.

We denote the background solution with a suffix 0.
In terms of the metric variables defined in (\ref{1}), we find
\be
a_0=1,\: b_0=S(t),\:  R_0=rS(t).
\ee
The background Hubble parameter is
\begin{equation}
H_0(t)=\fr{\dot{R_0}}{a_0R_0}=\frac{\dot{S}}{S}=\frac{\alpha}{t},
\end{equation}
and the energy density is calculated from the Friedmann equation,
\begin{equation}
\rho_0(t)=\frac{3\alpha^2}{8\pi Gt^2}.
\end{equation}
The energy density perturbation is defined as
\be
\de(t,r)\equiv \fr{\rho(t,r)-\rho_0(t)}{\rho_0(t)}.\label{defofdelta}%=\ti{\rho}(t,r)-1.\label{defofdelta}
\ee
We introduce a variable $H$ defined by
\begin{equation}
H(t,r)\equiv \fr{\dot{R}}{aR}=\fr{U}{R}.\label{defH}
\end{equation}
The curvature profile $K(t,r)$ is defined by rewriting $b$ as
\be
b(t,r)=\frac{R'(t,r)}{\sqrt{1-K(t,r)r^2}}\label{16}.
\ee
This quantity $K(t,r)$ vanishes outside the perturbed region so that the solution asymptotically approaches the background FLRW
 solution at spatial infinity.

We denote the comoving radius of a perturbed region by $r_\mathrm{i}$, the 
precise definition of which will be given later (see eq. (\ref{defri})), 
 and define a dimensionless parameter
$\ep$ in terms of the square ratio of the Hubble radius  $H_0^{-1}$ to the physical
length scale of the configuration,
\be
\ep \equiv \left(\frac{H_0^{-1}}{S(t)r_\mathrm{i}}\right)^2
=(\dot{S}r_\mathrm{i})^{-2}
=\frac{t_\mathrm{i}^{2\al}t^{\beta}}{\al^2 r_\mathrm{i}^2},\quad \beta\equiv 2(1-\al).\label{defofep}
\ee
When we set the initial conditions for PBH
formation, the size of the perturbed 
region is much larger than the
Hubble horizon. 
%This remains the case until the horizon mass becomes larger than
%the PBH mass. 
%The horizon mass grows with cosmic time
%(for the radiation-dominated regime,
%the growth is directly proportional to time).
This means $\ep\ll 1$  
at the beginning, so it can serve as an
expansion parameter to construct an analytic solution of 
the system of Einstein equations to describe the spatial dependence of
 all the above variables 
at the initial moment when we set the initial conditions. 
In this paper, the second order solution, obtained in PNY, is basically used to provide the initial conditions 
for the numerical computations. 

%For 
%the sake of brevity,  below we will call this dependence
%``time evolution''. 

We define the initial curvature profile as
%We write the initial condition for (\ref{31}) as
\be
K(0,r)\equiv K_\mathrm{i}(r),
\ee
where $K_\mathrm{i}(r)$ is an arbitrary function of $r$ 
which vanishes outside the perturbed region.
%
%and define a new variable $\tilde{K}$ by
%\be
%1-K(t,r)r^2=(1-K_\mathrm{i}(r)r^2)\tilde{K}(t,r).\label{34}
%\ee
%$\tilde{K}$ is unity at spatial infinity like the other tilde-variables, but
%in contrast to the other tilde-variables,
%it describes the evolution of curvature deviation from the initial curvature
%profile rather than the deviation from the spatially flat Friedmann universe. 
%
Note that,
from (\ref{16}), $K_\mathrm{i}(r)$ has to satisfy the condition
\be
K_\mathrm{i}(r)<\fr{1}{r^2}.\label{condforKi}
\ee
We normalize radial Lagrangian coordinate $r$ in such a way 
that $K_\mathrm{i}(0)=1$.

In order to represent the comoving length scale of the 
perturbed region,
we use the co-moving radius,
$r_\mathrm{i}$, of the overdense region.
We can calculate $r_{\mathrm i}$ by %defining energy density perturbation as
%\be
%\delta(t,r)\equiv\fr{\rho(t,r)-\rho_0(t)}{\rho_0(t)}=\ti{\rho}(t,r)-1
%\ee
%and 
solving the following equation for the energy density perturbation defined by (\ref{defofdelta}):
\be
\delta(t,r_{\mathrm i})=0.\label{defri}
\ee
Since the initial condition is taken at the superhorizon regime, when
$\epsilon$ is extremely small, the following lowest-order solution\cite{Polnarev:2006aa} %\t{ムスカを引用}
\be
\de(t,r)=\fr{2\ri^2}{9r^2}(r^3\Ki)'\ep(t)\label{firstorder}
\ee
suffices to calculate $r_\mathrm{i}$, which is obtained by solving 
\be
3K_\mathrm{i}(r_\mathrm{i})+r_{\mathrm i}K'_\mathrm{i}(r_\mathrm{i})=0. 
\label{rieq}
\ee
Note that the physical length scale in the
asymptotic Friedmann region
is obtained by multiplying by the scale factor $S(t)$, the 
normalization of which
we have not specified.  We can therefore  set up initial 
conditions for the PBH formation with arbitrary mass scales by adjusting
the normalization of $S(t)$ which appears in the expansion parameter.

We also introduce the averaged over-density, denoted by $\bar{\delta}$
and defined as the energy density perturbation averaged over the
over-dense region as follows:
\be
\bar{\delta}(t)\equiv\left(\fr{4}{3}\pi R(t,r_\mathrm{od}(t))^3\right)^{-1}\int^{R(t,r_\mathrm{od}(t))}_04\pi\delta R^2dR.\label{defdeltabar}
\ee
Here $r_\mathrm{od}(t)$ represents the comoving radius of the 
overdense region and is the solution of $\delta(t,r_{\mathrm{od}}(t))=0$.
It turns out that $r_\mathrm{od}(t)$ is very close to $r_\mathrm{i}$
calculated from (\ref{rieq}), i.e. lowest-order expansion.
 
Hereafter, 
we will always use the initial conditions obtained in this section.

%%%%%%%%%%%%%%%%%%%%%%%%%%%%%%%%%%%%%%%%%%%%%%%%%%%%%%%%%%%%%%%%%%
\section{Basic equations used in the numerical computations}
%%%%%%%%%%%%%%%%%%%%%%%%%%%%%%%%%%%%%%%%%%%%%%%%%%%%%%%%%%%%%%%%%%%
The following equations were used in
\cite{MayWhite} to analyze the gravitational collapse of spherically symmetric masses:
\be
\dot{U}=-a\left(4\pi R^2\fr{\Ga}{w}P'+\fr{MG}{R^2}
+4\pi GPR \right)\label{u_t},
\ee
\be
\dot{R}=aU,\label{R_t}
\ee
\be
\fr{(\nu R^2)^{\cdot}}{\nu R^2}=-a\fr{U'}{R'}\label{rhoRsq},
\ee
\be
\dot{E}=-P\left(\fr{1}{\nu}\right)^{\cdot}\label{E_t},
\ee
\be
\fr{(aw)'}{aw}=\fr{E'+P(1/\nu)'}{w},\label{aw}
\ee
\be
M=4\pi\int^r_0\rho R^2R' dr, \label{mdef}
\ee
\be
\Ga=4\pi \nu R^2R',\label{Gamma}
\ee
\be
P=\ga \rho,\label{eos}
\ee
\be
w=E+P/\nu,\label{w}
\ee
where $E\equiv \rho/\nu$ and
\be
\nu\equiv\fr{1}{4\pi bR^2}\label{rhob}.
\ee
The constraint equation reads
\be
\left(\fr{R'}{b}\right)^2=\Ga^2=1+U^2-\fr{2M}{R}.\label{constraint}
\ee
We rewrite these equations 
in terms of the barred variables, 
which are defined by factoring out 
the scale factor $S$ and the background energy density $\rho_0$ 
as shown below: 
%\be
%R=R_0\ti{R}=rS\ti{R}\equiv S\bar{R},
%\ee
\be
\bar{R}\equiv R/S,
\ee
%\be
%\ti{a}=a=\bar{a},
%\ee
\be
\bar{a}\equiv a,
\ee
%\be
%u=u_0\ti{u}=\dot{S}r\ti{u}\equiv \dot{S}\bar{u},
%\ee
\be
\bar{U}\equiv U/\dot{S},
\ee
%\be
%e=\rho_0\bar{\rho},
%\ee
\be
\bar{\rho}\equiv \rho/\rho_0,
\ee
%\be
%m=\fr{4}{3}\pi \rho_0R^3\ti{\mu}=\fr{4}{3}\pi \rho_0S^3\bar{R}^3\ti{\mu}\equiv \rho_0S^3\bar{M},
%\ee
\be
\bar{M}\equiv M/(\rho_0S^3),
\ee
%\be
%b=\fr{R'}{\sqrt{1-Kr^2}}=\fr{S\bar{R}'}{\sqrt{1-Kr^2}}\equiv S\bar{b},
%\ee
\be
\bar{b}\equiv b/S,
\ee
%\be
%\nu=\fr{1}{4\pi R^2b}=\fr{1}{4\pi S^3\bar{R}^2\bar{b}}\equiv S^{-3}\bar{\nu},
%\ee
\be
\bar{\nu}\equiv S^3\nu,
\ee
%\be
%\Ga=4\pi \nu R^2R_\mu=4\pi S^{-3}\bar{\nu} S^3\bar{R}^2\bar{R}_\mu=4\pi\bar{\nu}\bar{R}^2\bar{R}_\mu\equiv \bar{\Ga},
%\ee
\be
\bar{\Gamma}\equiv \Gamma,
\ee
%\be
%P\equiv \rho_0\bar{P},
%\ee
\be
\bar{P}\equiv P/\rho_0,
\ee
%\be
%w=E+\fr{P}{\nu}=\fr{e}{\nu}+\fr{P}{\nu}=\fr{\rho_0\bar{\rho}}{S^{-3}\bar{\nu}}+\fr{\rho_0\bar{P}}{S^{-3}\bar{\nu}}=\rho_0S^3\left(\fr{\bar{\rho}}{\bar{\nu}}+\fr{\bar{P}}{\bar{\nu}}\right)\equiv \rho_0S^3\bar{w}.
%\ee
\be
\bar{w}\equiv w/(\rho_0S^3).
\ee
First, using these definitions, (\ref{u_t}) leads to%にバー変数の定義を用いて
%をバー変数を用いて書き換えると、
%\be
%u_t=-a\left(4\pi R^2\fr{\Ga}{w}P_\mu+\fr{mG}{R^2}+4\pi GPR \right),
%\ee
%\be
%(\dot{S}\bar{U})^\cdot=-\bar{a}\left(4\pi S^2\bar{R}^2\fr{\bar{\Ga}}{\rho_0S^3\bar{w}}\rho_0\bar{P}_\mu+\fr{S^3\rho_0\bar{M}G}{S^2\bar{R}^2}+4\pi G\rho_0\bar{P}S\bar{R}\right),
%\ee
\be
\ddot{S}\bar{U}+\dot{S}\dot{\bar{U}}
=-\bar{a}\left(4\pi S^2\bar{R}^2\fr{\bar{\Ga}}{S^3\bar{w}}\bar{P}'
+\fr{S\rho_0\bar{M}G}{\bar{R}^2}+4\pi G\rho_0\bar{P}S\bar{R}\right).\label{ubar}
\ee
Using the Friedmann equation
\be
\left(\fr{\dot{S}}{S}\right)^2=\fr{8\pi G}{3}\rho_0\label{frie}
\ee
on the right side of (\ref{ubar}) we find
\be
\ddot{S}\bar{U}+\dot{S}\dot{\bar{U}}
=-\bar{a}\left(4\pi S^2\bar{R}^2\fr{\bar{\Ga}}{S^3\bar{w}}\bar{P}'
+\fr{3}{8\pi}\fr{\dot{S}^2}{S}\fr{\bar{M}}{\bar{R}^2}
+\fr{3}{2}\fr{\dot{S}^2}{S}\bar{P}\bar{R}\right).
\ee
Dividing the both sides by $\dot{S}^2/S$ and noting
\be
S=\left(\fr{t}{t_i}\right)^\al,\quad \fr{S\ddot{S}}{\dot{S}^2}=\fr{\al-1}{\al},
\ee
we obtain
\be
\fr{t}{\al}\dot{\bar{U}}
=-\bar{a}\left(4\pi \dot{S}^{-2}\bar{R}^2\fr{\bar{\Ga}}{\bar{w}}\bar{P}'
+\fr{3}{8\pi}\fr{\bar{M}}{\bar{R}^2}+\fr{3}{2}\bar{P}\bar{R}\right)+\fr{1-\al}{\al}\bar{U}.
\ee
%$\ep$の定義
%\be
%\ep\equiv (\dot{S}r_i)^{-2}=\fr{t_i^{2\al}t^\beta}{\al^2r_i^2},
%\ee
Defining a new time coordinate by
\be
\tau\equiv \fr{\al}{\beta}\log\ep\label{deftau}
\ee
and noting the following relationship
\be
\fr{t}{\al}\fr{\pa}{\pa t}=\fr{\pa}{\pa \tau},\label{ttotau}
\ee
which is obtained from the definition of $\ep$, (\ref{defofep}),
one finds
\be
(\bar{U})_\tau
=-\bar{a}\left(4\pi r_i^2\exp\left(\fr{\beta}{\al}\tau\right)\bar{R}^2\fr{\bar{\Ga}}{\bar{w}}\bar{P}'
+\fr{3}{8\pi}\fr{\bar{M}}{\bar{R}^2}+\fr{3}{2}\bar{P}\bar{R}\right)+\fr{1-\al}{\al}\bar{U},\label{Ubartau}
\ee
where the subscript $\tau$ denotes a time derivative with respect to $\tau$.
%%%%%%%%%%%%%%%%%%%%%%%%%%%%%%%%%%%%%%%%%%%%%%%%%%%%%%%%%%%%%%%%%%%%%%%%%%%%%%%%%%
Similarly, (\ref{R_t})-(\ref{E_t}) are rewritten in terms of the 
barred variables as follows:
%\be
%R_t=au,
%\ee
%\be
%\dot{S}\bar{R}+S\dot{\bar{R}}=\bar{a}\dot{S}\bar{U},
%\ee
%\be
%\fr{t}{\al}\dot{\bar{R}}=\bar{a}\bar{U}-\bar{R},
%\ee
\be
\bar{R}_\tau=\bar{a}\bar{U}-\bar{R},\label{Rbartau}
\ee
%\be
%\fr{(\nu R^2)_t}{\nu R^2}=-a\fr{u_\mu}{R_\mu},
%\ee
%\be
%\fr{(S^{-3}\bar{\nu}S^2\bar{R}^2)_t}{S^{-3}\bar{\nu}S^2\bar{R}^2}
%=\fr{(S^{-1})^\cdot\bar{\nu}\bar{R}^2+S^{-1}(\bar{\nu}\bar{R}^2)_t}{S^{-1}\bar{\nu}\bar{R}^2}
%=-\bar{a}\fr{\dot{S}\bar{U}_\mu}{S\bar{R}_\mu},
%\ee
%\be
%-\fr{\dot{S}}{S}+\fr{(\bar{\nu}\bar{R}^2)_t}{\bar{\nu}\bar{R}^2}
%=-\bar{a}\fr{\dot{S}}{S}\fr{\bar{U}_\mu}{\bar{R}_\mu},
%\ee
%\be
%\fr{t}{\al}\fr{(\bar{\nu}\bar{R}^2)_t}{\bar{\nu}\bar{R}^2}=-\bar{a}\fr{\bar{U}_\mu}{\bar{R}_\mu}+1,
%\ee
\be
\fr{(\bar{\nu}\bar{R}^2)_\tau}{\bar{\nu}\bar{R}^2}
=-\bar{a}\fr{\bar{U}'}{\bar{R}'}+1,\label{nubartau}
\ee
%%%%%%%%%%%%%%%%%%%%%%%%%%%%%%%%%%%%%%%%
%\be
%E_t=-P\left(\fr{1}{\nu}\right)_t,
%\ee
%\be
%(S^3\rho_0\bar{\rho})_t
%=-\rho_0\bar{P}\left(\fr{1}{S^{-3}\bar{\nu}}\right)_t
%=-\rho_0\bar{P}\left(\left(\fr{1}{S^{-3}}\right)_t\fr{1}{\bar{\nu}}+\fr{1}{S^{-3}}\left(\fr{1}{\bar{\nu}}\right)_t\right),
%\ee
%\be
%\fr{(S^3\rho_0)_t}{S^3\rho_0}\bar{\rho}+\bar{\rho}_t
%=-\bar{P}\left(\fr{1}{S^3}\left(\fr{1}{S^{-3}}\right)_t\fr{1}{\bar{\nu}}+\left(\fr{1}{\bar{\nu}}\right)_t\right),
%\ee
%\be
%\left(\fr{3\al}{t}-\fr{2}{t}\right)\bar{E}+\bar{E}_t
%=-\bar{P}\left(\fr{3\al}{t}\fr{1}{\bar{\nu}}+\left(\fr{1}{\bar{\nu}}\right)_t\right),
%\ee
%\be
%\left(3-\fr{2}{\al}\right)\bar{E}+\bar{E}_\tau=-\bar{P}\left(\fr{3}{\bar{\nu}}+\left(\fr{1}{\nu}\right)_\tau\right),
%\ee
\be
\bar{E}_\tau
=-\bar{P}\left(\fr{1}{\bar{\nu}}\right)_\tau.\label{Ebartau}
\ee
%%%%%%%%%%%%%%%%%%%%%%%%%
%\subsection{equation(5)}
%\be
%\fr{(aw)_\mu}{aw}=\fr{E_\mu+P(1/\nu)_\mu}{w},
%\ee
%this equation does not contain any time derivatives, 
%so the new equation takes the same form as the equation above:
Since the barred variables are defined by factoring out 
$S$ and $\rho_0$, 
depending only on time, 
the equations (\ref{aw})-(\ref{rhob}), which do not contain time derivatives, 
do not change their appearance even after being rewritten in terms of the barred variables:
\be
\fr{(\bar{a}\bar{w})'}{\bar{a}\bar{w}}=\fr{\bar{E}'+\bar{P}(1/\bar{\nu})'}{\bar{w}},\label{abarwbar}
\ee
\be
\bar{M}=4\pi\int^r_0\bar{\rho}\bar{R}^2\bar{R}' dr,\label{Mbar}
\ee
\be
\bar{\Ga}=4\pi \bar{\nu} \bar{R}^2\bar{R}',\label{Gabar1}
\ee
\be
\bar{P}=\ga\bar{\rho},\label{Pbar}
\ee
\be
\bar{w}=\bar{E}+\bar{P}/\bar{\nu},\label{wbar}
\ee
\begin{equation}
\bar{\nu}=\frac{1}{4\pi \bar{b}\bar{R}^2}.\label{nubar}
\end{equation}
%\be
%\fr{(\bar{a}\bar{w})'}{\bar{a}\bar{w}}=\fr{\bar{E}'+\bar{P}(1/\bar{\nu})'}{\bar{w}},\label{aderi}
%\ee
%The following equation, which is equivalent to this equation, 
%may be useful.
%%%%%%%%%%%%%%%%%%
%\subsection{equation(6)}
%\be
%m=4\pi\int^\mu_0eR^2R_\mu d\mu,
%\ee
%\be
%\bar{M}=4\pi\int^r_0\bar{\rho}\bar{R}^2\bar{R}' dr,
%\ee
%%%%%%%%%%%%%
%\subsection{equation(7)}
%\be
%\Ga=4\pi \nu R^2R_\mu,
%\ee
%\be
%\bar{\Ga}=4\pi \bar{\nu} \bar{R}^2\bar{R}',
%\ee
%%%%%%%%%%%%%%
%\subsection{equation(8)}
%\be
%P=\ga e,
%\ee
%\be
%\rho_0\bar{P}=\ga \rho_0\bar{\rho},
%\ee
%\be
%\bar{P}=\ga\bar{\rho},
%\ee
%%%%%%%%%%%
%\subsection{equation(9)}
%\be
%w=E+P/\nu,
%\ee
%\be
%\rho_0S^3\bar{w}=\rho_0S^3\bar{E}+\rho_0\bar{P}/(S^{-3}\bar{\nu}),
%\ee
%\be
%\bar{w}=\bar{E}+\bar{P}/\bar{\nu},
%\ee
%\be
%\bar{\nu}=\fr{1}{4\pi \bar{b}\bar{R}^2}.\label{rhobarbbar}
%\ee
The following equation is equivalent to (\ref{abarwbar})
and is also useful:
\be
\bar{a}=\bar{\rho}^{-\fr{\ga}{1+\ga}}.\label{abar}
\ee
Lastly, 
the constraint equation (\ref{constraint}) can be rewritten as
\be
\bar{\Ga}=1+\dot{S}^2(\bar{U}^2-\fr{3\bar{M}}{4\pi \bar{R}}).\label{constraintbar}
\ee
This equation is not used for time evolution
but rather is used to estimate numerical errors. 
%%%%%%%%%%%%%
Now, all the relevant equations in terms of the barred variables, 
(\ref{Ubartau}), (\ref{Rbartau}), (\ref{nubartau}), (\ref{Ebartau}),  (\ref{Mbar}), (\ref{Gabar1}), (\ref{Pbar}), (\ref{wbar}), (\ref{nubar}), (\ref{abar}) and (\ref{constraintbar})
have been obtained.
\begin{comment}
\begin{equation*}
(\bar{U})_\tau
=-\bar{a}\left[4\pi r_i^2\exp\left(\fr{\beta}{\al}\tau\right)\bar{R}^2\fr{\bar{\Ga}}{\bar{w}}\bar{P}'
+\fr{3}{8\pi}\fr{\bar{M}}{\bar{R}^2}+\fr{3}{2}\bar{P}\bar{R}\right]+\fr{1-\al}{\al}\bar{U},
%\label{sumarryini}
\end{equation*}
\begin{equation*}
\bar{R}_\tau=\bar{a}\bar{U}-\bar{R},
\end{equation*}
\begin{equation*}
\fr{(\bar{\nu}\bar{R}^2)_\tau}{\bar{\nu}\bar{R}^2}=-\bar{a}\fr{\bar{U}'}{\bar{R}'}+1,
\end{equation*}
\begin{equation*}
\bar{E}_\tau
=-\bar{P}\left(\fr{1}{\bar{\nu}}\right)_\tau,
\end{equation*}
\begin{equation*}
\bar{a}=\bar{\rho}^{-\fr{\ga}{1+\ga}},
\end{equation*}
\begin{equation*}
\bar{M}=4\pi\int^r_0\bar{\rho}\bar{R}^2\bar{R}' dr,
\end{equation*}
\begin{equation*}
\bar{\Ga}=4\pi \bar{\nu} \bar{R}^2\bar{R}',
\end{equation*}
\begin{equation*}
\bar{P}=\ga\bar{\rho},
\end{equation*}
\begin{equation*}
\bar{w}=\bar{E}+\bar{P}/\bar{\nu},
\end{equation*}
\begin{equation*}
\bar{\Ga}=1+\dot{S}^2(\bar{U}^2-\fr{3\bar{M}}{4\pi \bar{R}}).
%\label{sumarrylast}
\end{equation*}
\end{comment}
The boundary conditions are imposed such that
$\bar{U}=\bar{R}=\bar{M}=0$ and $\Ga=1$ at the centre, 
and $\bar{a}=\bar{\rho}=1$ on the outer boudary
so that the numerical solution is smoothly connected to the 
FLRW solution.

In the next section we discuss the numerical solutions of the basic equations 
presented here.
%%%%%%%%%%%%%%%%%%%%%%%%%%%%%%%%%%%%%%%%%%%%%%%%
%%%%%%%%%%%%%%%%%%%%%%%%%%%%%%%%%%%%%
%%%%%%%%%%%%%%%%%%%%%%%%%%%%%%%%%%%%%%
%\section{Reviewing the time evolution of the perturbation in the cosmic time slicing}
\section{Typical time evolution of perturbed regions in the cosmic
 time slicing: A review}
%%%%%%%%%%%%%%%%%%%%%%%%%%%%%%%%%%%%%%
%%%%%%%%%%%%%%%%%%%%%%%%%%%%%%%%%%%%%%%
%%%%%%%%%%%%%%%%%%%%%%%%%%%%%%%%%%%%%%%%%%%%%%%%%

\begin{figure}[h]
\begin{center}
\includegraphics[width=12cm,keepaspectratio,clip]{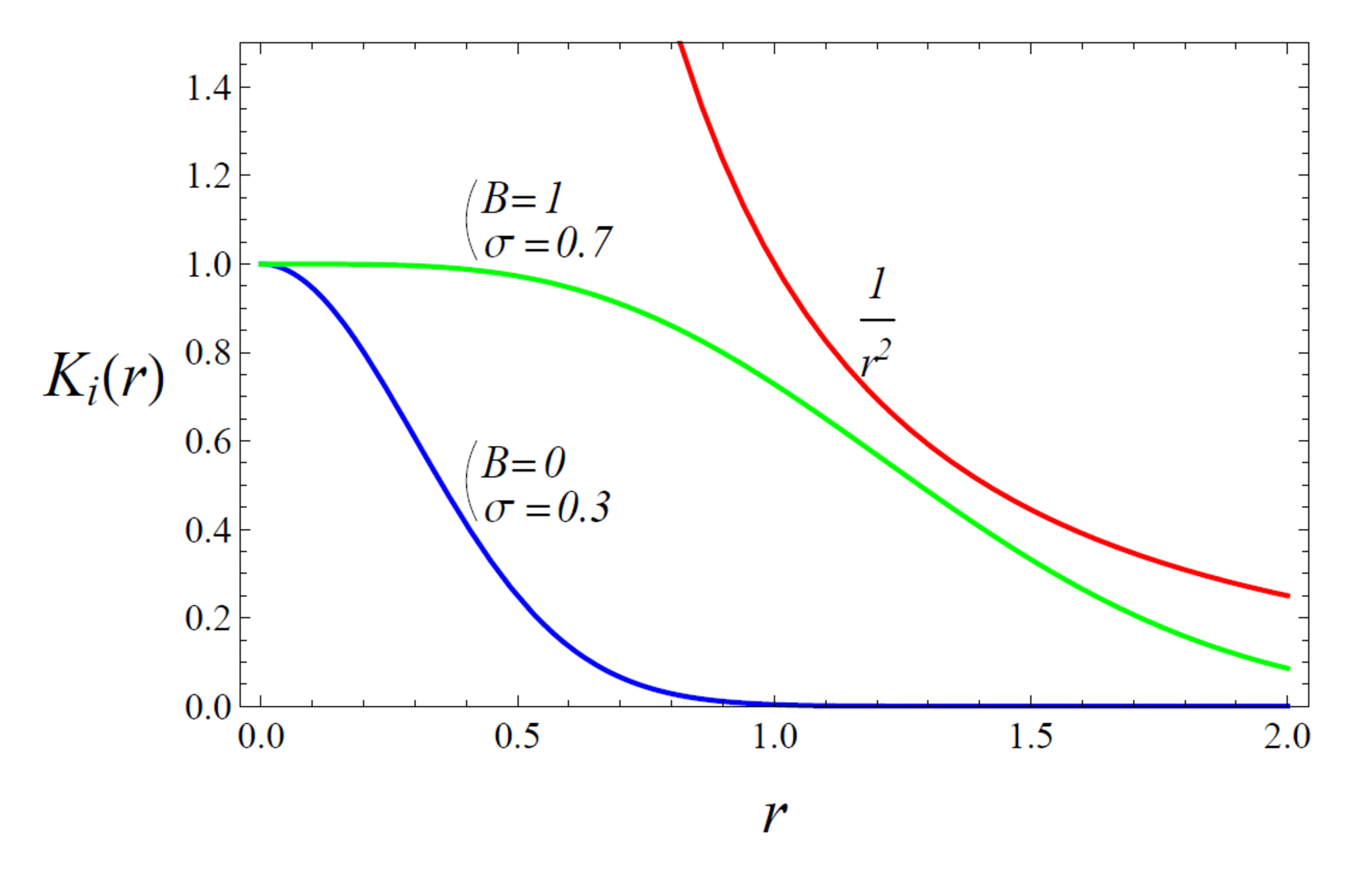}%jpgの方がmassを追加した
\end{center}
\caption{
A wide and narrow initial curvature profiles $\Ki$ represented by (\ref{original}). 
Note that $\Ki$ has to satisfy the condition $\Ki<1/r^2$. 
}
\label{originalprofile414}
\end{figure}

In order to test our numerical code which solves the evolution equations
described
in the previous section, we first adopt the
following two-parameter family of curvature profile
\be
\Ki=\left[1+\fr{B}{2}\left(\fr{r}{\si}\right)^2\right]\exp\left[-\fr{1}{2}
\left(\fr{r}{\si}\right)^2\right],\label{original}
\ee
which was also investigated in \cite{Polnarev:2006aa}(hereafter PM), in
order to examine the consistency with the results obtained in PM. 
As mentioned previously, the 
amplitude of the profile is set to unity at the origin, 
meaning that we use the same normalization as a spatially closed Friedmann universe. 
Here the parameters $B$ and $\si$ control the shapes of initial perturbations. 
The range of $B$ is limited to $0\leq B\leq 1$ so that $\Ki$ is a monotonic function. 
Two examples of (\ref{original}) are shown in Figure \ref{originalprofile414}. \\
%%%%%%%%%%%%%%%%%%%%%%%%%%%%%%%%%%%%%%%%%%%%%%%%%%%%%%%%%%%%%%%%%%%%%%%%%5
%%%%%%%%%%%%%%%%%%%%%%%%%%%%%%%%%%%%%%%%%%%%%%%%%%%%%%%%%%%%%%%%%%%%%%%%%%%

In order to relate the initial curvature perturbation to 
the amplitude of the density perturbation, 
following PM, let us approximately evaluate the energy density perturbation 
averaged over the over-dense region, denoted by $\bar{\de}$ and defined by (\ref{defdeltabar}), at the time of the horizon crossing. 
Using (\ref{firstorder}) $\bar{\de}(t)$ becomes
\be
\bar{\de}(t)=\left(\fr{4}{3}\pi \ri^3\right)^{-1}\int_0^{\ri}\fr{8\pi\ri^2}{9}(r^3\Ki)'\ep(t)dr=\fr{2}{3}K_\m{i}(r_\m{i})\ri^2\ep(t).\label{deltabarfirst}
\ee
By setting $\ep(t)=1$, we define
%In order to clarify the physical meaning of this condition, 
%following ムスカ let us introduce
\be
\bar{\de}_\m{hc}\equiv \fr{2}{3}K_\m{i}(r_\m{i})r_\m{i}^2.\label{dhc}
\ee
Note that (\ref{deltabarfirst}) is valid only when $\ep(t)\ll 1$,
so this formula gives just an approximate value of $\bar{\de}(t)$ at the time of the horizon crossing. 
Still, (\ref{dhc}) gives a good indicator of how strong gravity is. 
%This is an approximated value of $\bar{\delta}$, defined by (\ref{defdeltabar}), 
%which is evaluated at the time of horizon crossing of the perturbed region 
%by the first order expansion in $\ep$ with $\ep$ set to unity and 
%is an indication of how strong gravity is. 

A profile for which $\dhc$ is small corresponds to a small amplitude perturbation 
and PBH is not formed from this kind of initial configurations. 
%What matters is 
%the threshold value of $\dhc$, which is examined in the next section. 
The narrow profile in Figure \ref{originalprofile414} with $(B,\si)=(0,0.3)$ corresponds to $\dhc=0.04$, 
while the wide profile with $(B,\si)=(1,0.7)$ corresponds to $\dhc=0.51$. 
So the width of $\Ki$ approximately tells how large $\dhc$ is, as can be seen from (\ref{dhc}). 
In the following the time evolution of the initial perturbed region for these two cases is presented as illustration. 
%In PNY we presented the analytic solution obtained by the asymptotic expansion 
%We solve the recursive relations obtained in \S I\hspace{-.1em}V 
%for 
%four specific curvature profiles of the form
%\be
%K_\mathrm{i}(r)=\left[1+\fr{B}{2} \left(\frac{r}{\sigma}\right)^2\right]
%\exp\left[-\fr{1}{2}\left(\frac{r}{\sigma}\right)^2\right], \label{Kini}
%\ee
%where $B$ describes slope of
%curvature profiles and $\sigma$ specifies the comoving
%length scale of curvature profile.
%Smaller values of $B$ correspond to shallower profiles,
%and when $B=0$ the profile is simply Gaussian. 
  
%In the following, the time evolution of the perturbed regions are illustrated 
%for two cases with and without PBH formation. 

\begin{figure}[h]
\begin{center}
\includegraphics[width=14cm,keepaspectratio,clip]{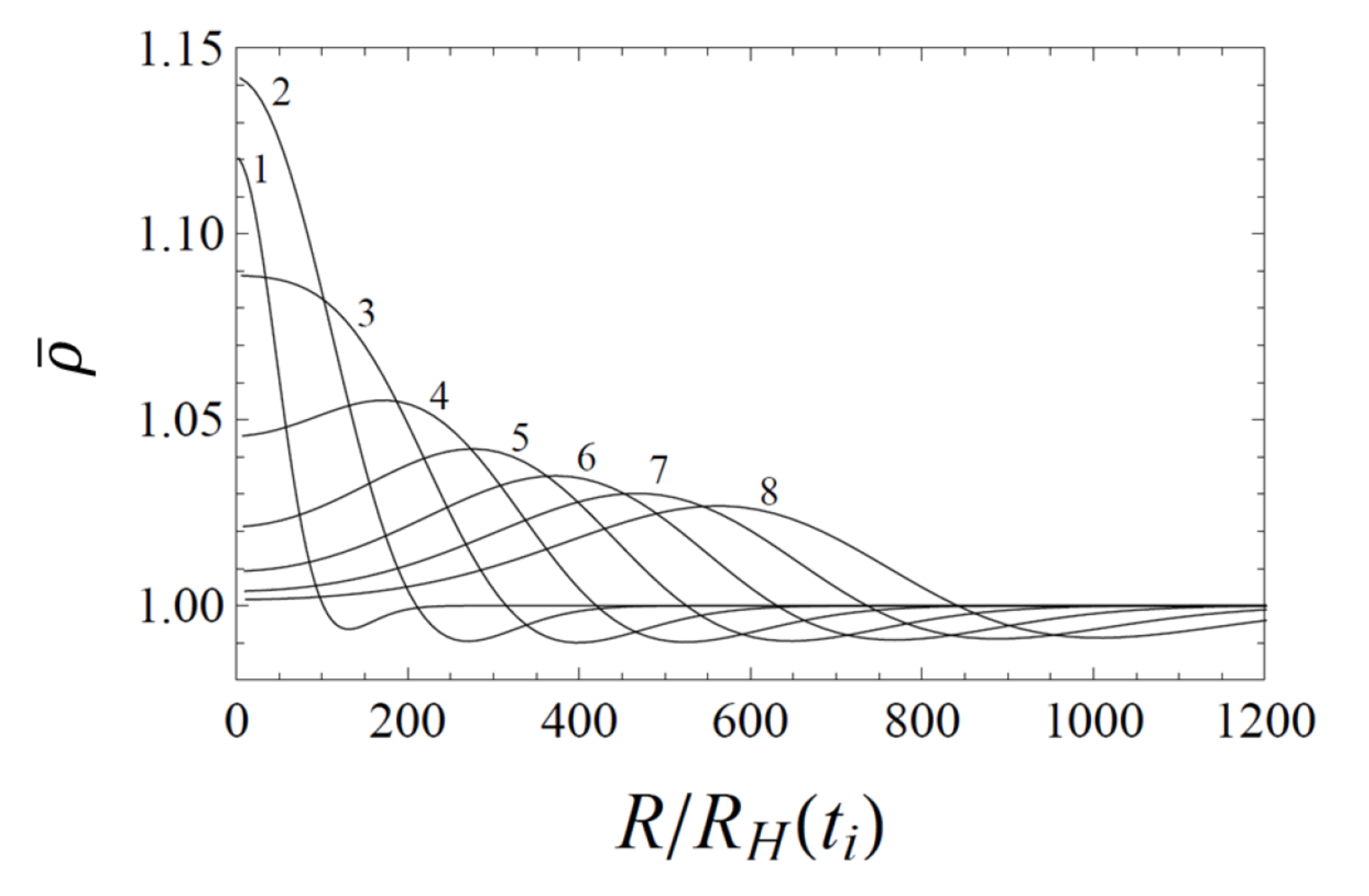}
\end{center}
\caption{
An example of the time evolution of the 
energy density perturbation $\bar{\rho}$, normalised by the background energy density, 
in a case where $(B,\sigma)=(0,0.3)$ and no PBH is formed. 
The plots are numbered in order of time evolution. 
}
\label{ebar_003}
\end{figure}

First, let us look at the case of the narrow profile ($\dhc=0.04$).
%First, let us look at a case 
%where the amplitude of the initial curvature perturbation is too small 
%for a PBH to be formed. 
The time evolution of the energy density $\bar{\rho}$, 
normalised by the background energy density, is shown
 in Figure  \ref{ebar_003}. 
%%%%%%%%%%%%%%%%%%%%%%%%%%%%%%%%%%%%%%%%%%%%%%%%%%%%%%%%%%%%%%%%%%%%%%%
%%%%%%%%%%%%%%%%%%%%%%%%%%%%%%%%%%%%%%%%%%%%%%%%%%%%%%%%%%%%%%%%%%%%%%%%
In this case 
the growth of the perturbation stops at some point in time after the horizon crossing 
and the energy density starts to decrease in the central region, with  
a sound wave propagating outward, the amplitude of which gradually decreases. 
That is, the initial perturbation dies away and the eventual state at $t=\infty$ is the 
flat FLRW universe. 

\begin{figure}[h]
\begin{center}
\includegraphics[width=14cm,keepaspectratio,clip]{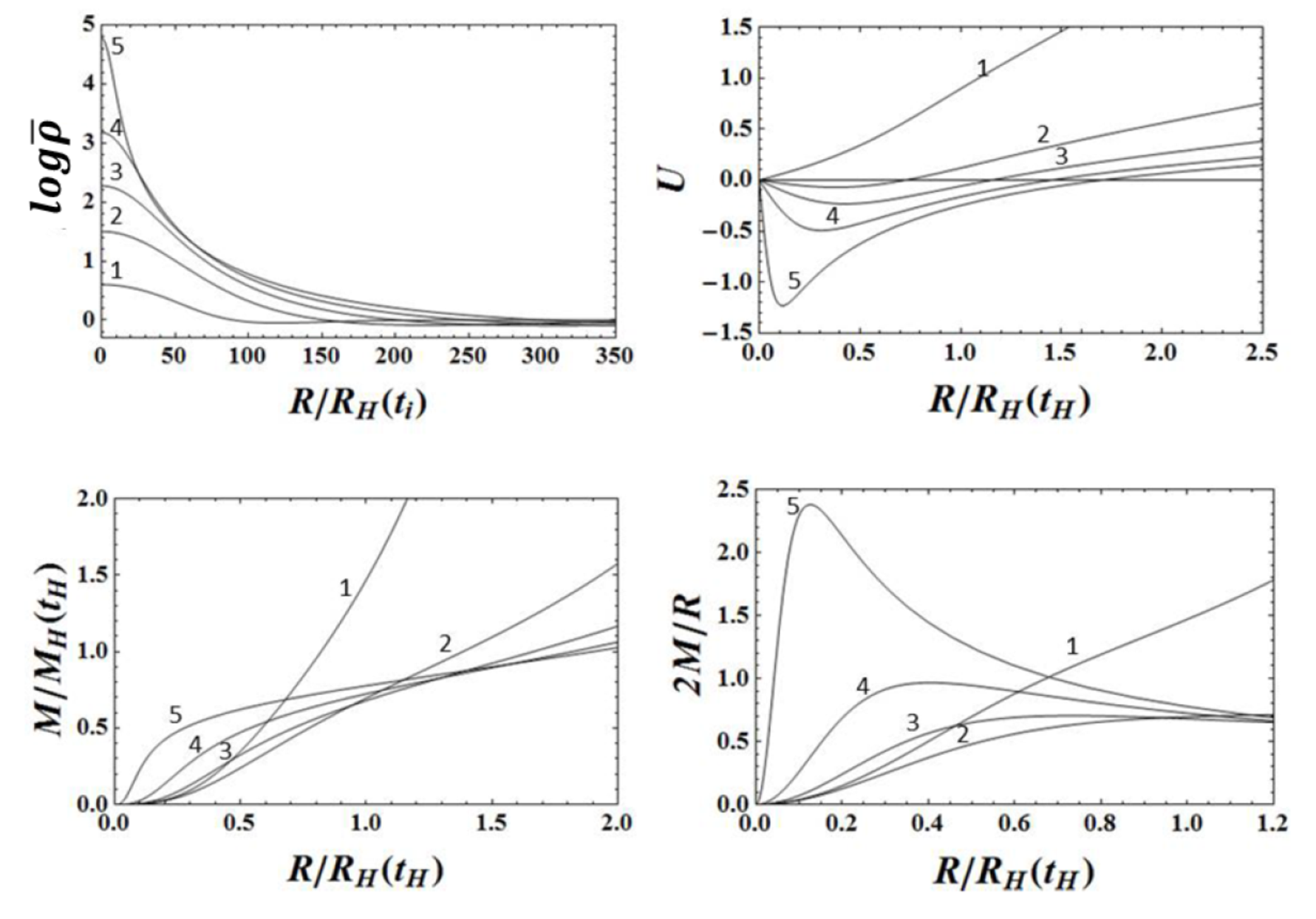}
\end{center}
\caption{
Examples of time evolution of $\bar{\rho}$(top-left), $U$(top-right), $M$(down-left) and $2M/R$(down-right), calculated using the cosmic time slicing. 
These are obtained for the case
$(B,\sigma)=(1,0.7)$ and a PBH is formed.
Each line is numbered in order of the time evolution. 
}
\label{cs_figures}
\end{figure}

Next, let us consider a case where the amplitude of the initial perturbation is 
sufficiently large ($(B,\si)=(1,0.7), \dhc=0.51$) and a PBH is eventually formed. 
The time evolution of $\bar{\rho}$, $U$ and $2M/R$ in this case is shown in Figure  \ref{cs_figures}.

%%%%%%%%%%%%%%%%%%%%%%%%%%%%%%%%%%%%%%%%%%%%%%%%%%%%%%%%%%%%%%%%%%%%%%%%%5
%%%%%%%%%%%%%%%%%%%%%%%%%%%%%%%%%%%%%%%%%%%%%%%%%%%%%%%%%%%%%%%%%%%%%%%%%%%
Near the centre the energy density increases drastically 
and the central perturbed region gradually expands outward. The central perturbed region is always
surrounded by the under-dense region. 
%the energy density is low in the surrounding region. 

From (\ref{defH}) $U$ is written as $U=HR$ 
and it corresponds to the recession velocity in the FLRW universe. 
At an early stage, when the amplitude of the perturbation is small,
$U$ is positive everywhere-reflecting the expansion of the universe. 
In the central region where gravity becomes increasingly stronger, 
the expansion decelerates rapidly and therefore $U$ decreases rapidly. 
Then at some point in time, 
the central region stops expanding 
%the expansion stops in the central region 
and 
$U$ becomes negative there, starting
gravitational contraction.

The mass $M$ is defined by (\ref{defofM}) 
and represents the total energy contained in a sphere of radius $R$. 
When the amplitude of perturbation is small, 
$M\sim \rho_0(t)R(t,r)^3$ hence 
$2M/R\propto R^2$. 
As mentioned earlier, 
at a later stage the energy density 
in the central region increases dramatically 
as a black hole is formed, 
which is contrasted with the decrease in the energy density 
in the surrounding region due to the expansion of the universe. 
%In contrast, the energy density decreases 
%in the surrounding region as the universe expands. 
In such a situation 
the mass profile in the central region is steep but is relatively gentle in the surrounding region. 
This feature of the mass profile can be easily understood 
by an analogy with a situation where a star resides in the vacuum, 
in which case the mass profile is a monotonically increasing function inside the radius of the star 
and is flat outside that radius. 
Due to this behaviour of $M$ at a later stage, 
in the region away from the centre 
the mass only weakly depends on $R$ and so approximately 
$2M/R\propto R^{-1}$ there. 
This means that 
a peak appears in the profile of $2M/R$ at some moment 
in cases where perturbation grows sufficiently. 
Specifically, in cases where a black hole is formed, 
the height of this peak exceeds unity and 
this implies the formation of the apparent horizon 
from the following arguments\cite{CriticalMusco}. 

Suppose the trajectory of a photon moving outward, along which 
\be
adt=bdr,
\ee
so along the geodesic of this photon,
\be
\fr{dR}{dt}=\fr{\pa R}{\pa t}dt+\fr{\pa R}{\pa r}dr=a(U+\Ga)dt,
\ee
where the definition of $U$ (\ref{2}) and $\Ga=R'/b$, 
followed from (\ref{Gamma}) and (\ref{rhob}), have been used. 
From this we find $dR/dt=0$ where $U=-\Ga$ holds, 
meaning the photon reaching that point cannot 
escape further away from the centre. 
Since we find $U=-\Ga$ when $2M/R=1$ from the constraint (\ref{constraint}), 
the peak of $2M/R$ with the height exceeding unity means that there exist 
photons which are trapped by the gravitational potential of the central black hole 
and cannot escape to infinity. Namely, the apparent horizon has been formed. %meaning the formation of a horizon. 

We have calculated the time evolution with various values of $B$ and
$\sigma$
and obtained results fully consistent with PM. 
In the next section, we discuss the PBH formation condition for 
a much wider class of profiles than that defined by (\ref{original})
.

%On the other hand, from the constraint equation (\ref{constraint}), 
%$2M/R=1$ when $U=-\Ga$. 
\section{Two master parameters cruicial for PBHs formation }
\begin{figure}[h]
\begin{center}
\includegraphics[width=16cm,keepaspectratio,clip]{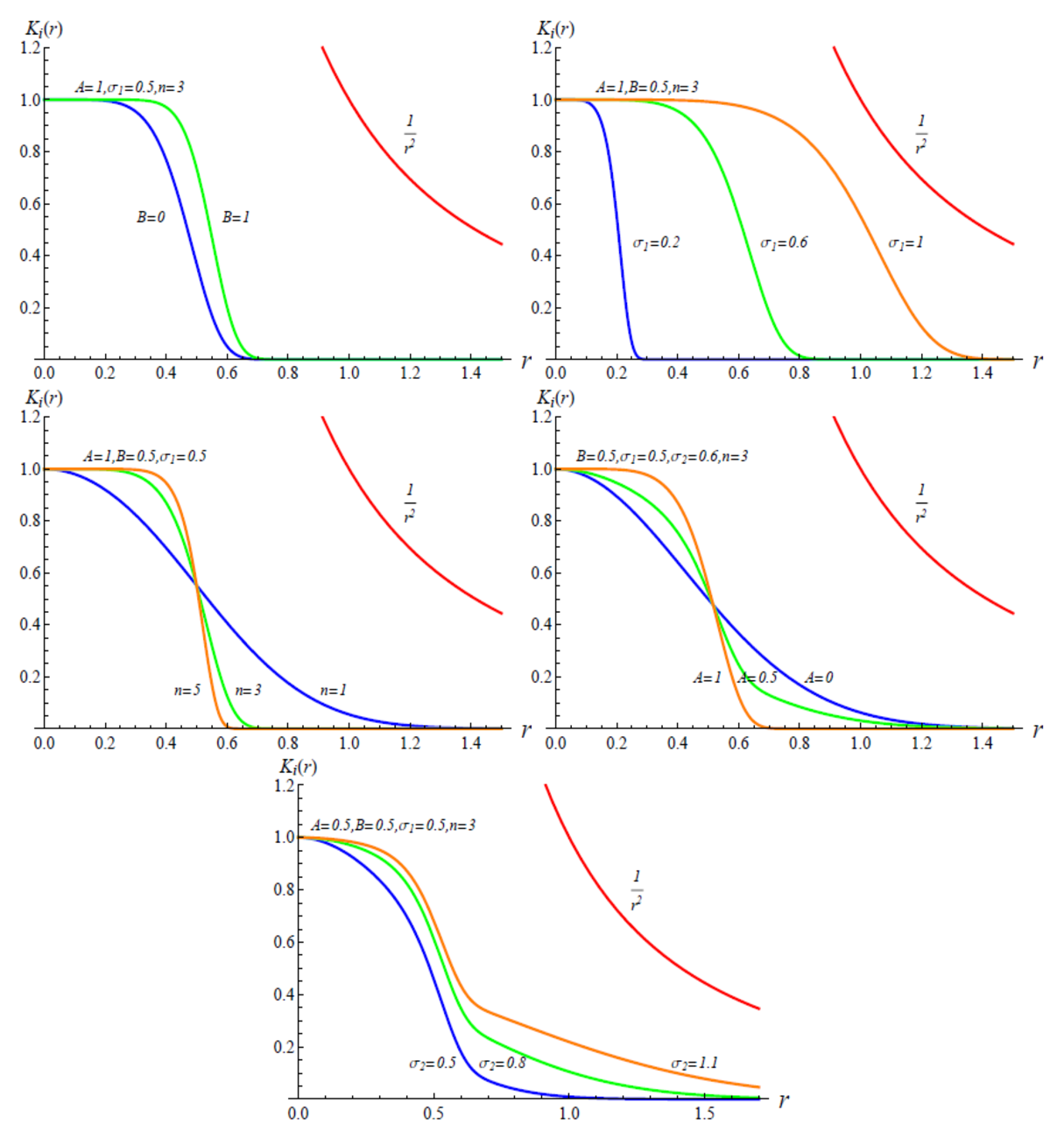}%jpgの方がmassを追加した
\end{center}
\caption{
Dependence of the shapes of profiles represented by (\ref{newfunction}) on parameters. 
In each panel, the dependence of the shape on one of the parameters is shown with the rest of the parameters fixed. 
%Two profiles with $B=0$ and $B=1$ are shown in the left panel, where $\sigma$ and $n$ are the same ($\si=0.5, n=3$). 
%This shows increasing $B$ makes profile slightly wider and steeper. 
%The central and right panel tell that increasing $\si$ makes profile wider while a larger value of $n$ corresponds to a steeper profile. 
}
\label{shapes608}
\end{figure}

We now proceed to our full analysis
introducing the following function
\be
K_\mathrm{i}(r)=A\left[1+B\left(\frac{r}{\sigma_1}\right)^{2n}\right]
\exp\left[-\left(\fr{r}{\sigma_1}\right)^{2n}\right]
+(1-A)\exp\left[-\left(\fr{r}{\sigma_2}\right)^{2}\right],\label{newfunction}
\ee
which can represent various shapes of profiles using the five parameters
as is shown in Figure \ref{shapes608}. 
Profiles with $A=1$ include the profiles studied in PM but can
also incorporate those with a steeper transition to the homogeneous region,
%are equivalent to or quite similar to those investigated by two functions 
 whereas the inclusion of the second term makes it possible to further 
realize  profiles with tails (see Figure \ref{shapes608}). 
It also turned out that this function with $n=2$
can fit the profiles investigated in \cite{Shibata:1999zs} sufficiently well, 
after translating their profiles into our $K_\m{i}(r)$ since a different coordinate system was used their.  
Thus our function not only includes those investigated in previous work but also 
enables us to investigate new shapes of profiles. 

\begin{figure}[h]
\begin{center}
\includegraphics[width=14cm,keepaspectratio,clip]{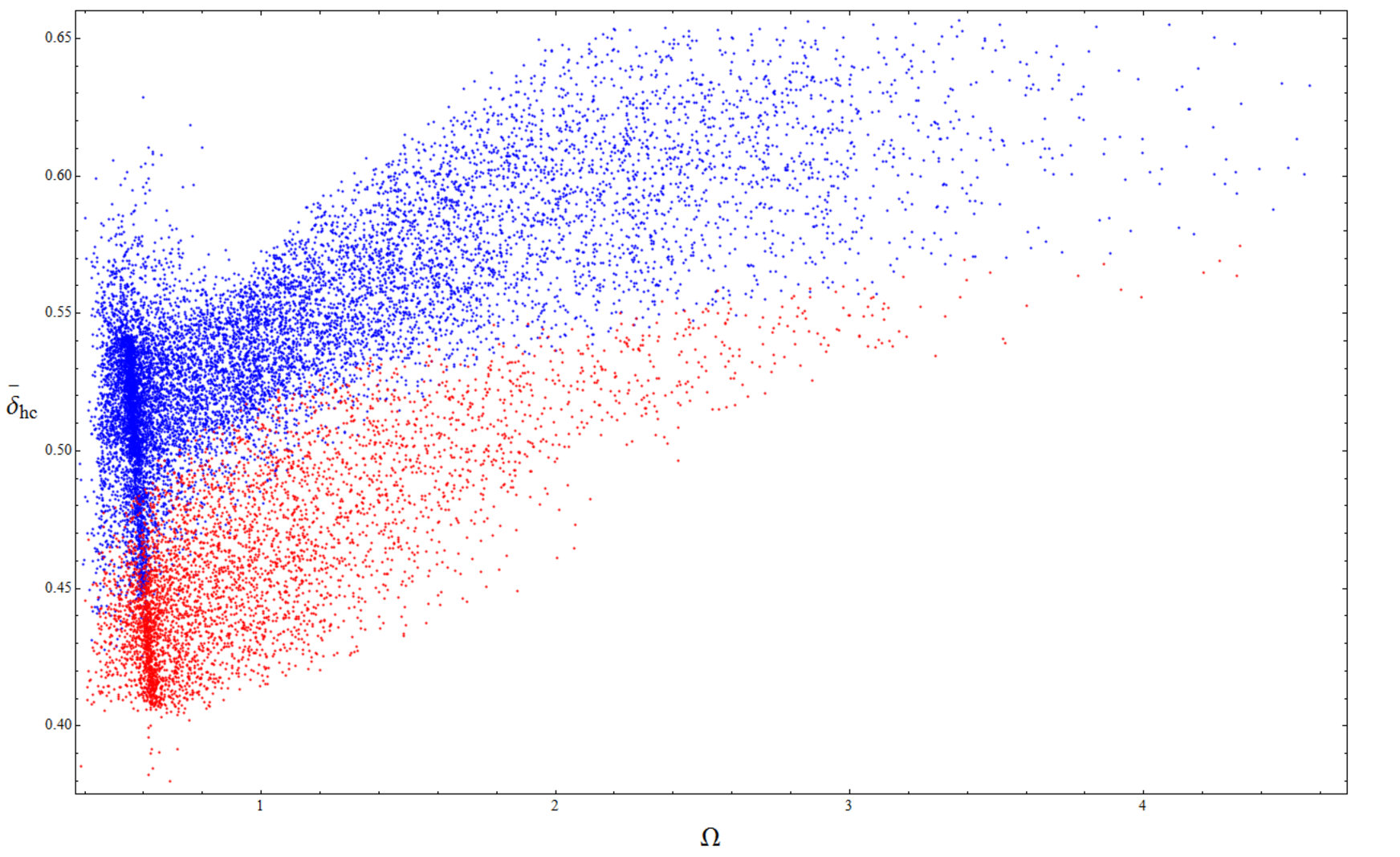}
\end{center}
\caption{
The blue points correspond to profiles which lead to PBH formation 
and the red points to profiles which do not. Blue and red points are not well
separated by a single line in this diagram.
}
\label{fail531}
\end{figure}

For each set of five parameters, the time evolution of the perturbation 
is calculated to reveal whether or not a PBH is formed at the end. 
%%%%%%%%%%%%%%%%%%%%%%%%%%%%%%%%%%%%%%%%%%%%%%%%%%%%%%%%%%%%%%%%%%%%%%%%%5
%%%%%%%%%%%%%%%%%%%%%%%%%%%%%%%%%%%%%%%%%%%%%%%%%%%%%%%%%%%%%%%%%%%%%%%%%%%
First, let us try to describe PBH formation condition for these various kinds of profiles using the quantity $\dhc$
and 
\be
\Omega\equiv \max_r|K_\m{i}'(r)|,
\ee
which corresponds to $\Delta$ in PM and may provide a measure of the
density gradient, or the pressure gradient for the case of the radiation-dominated universe.
As can be seen in Figure  \ref{fail531}, these two quantities fail to 
distinguish between the profiles
which lead to PBH formation and those which do not. 
%%%%%%%%%%%%%%%%%%%%%%%%%%%%%%%%
%%%%%%%%%%%%%%%%%%%%%%%%%%%%%%%%%%%%%
One conceivable reason why this combination fails is because $\dhc$ is too sensitive to the edge $r=r_\m{i}$ of 
configurations. To see this let us consider two profiles with almost the same shape near the centre but one of these has 
slightly wider tail with a larger $r_\m{i}$ than the other. 
For this wider profile, $\dhc$ is smaller since this quantity is the energy density perturbation 
averaged inside $r<r_\m{i}$. 
However, what is important in physically determining whether a PBH is formed or not is the shape of profile near the centre, while 
the information around the outer boundary is 
 less important. Therefore, proper quantities which can 
clearly distinguish between the PBH-forming profiles and the non-forming ones should take the same value for those two profiles with 
the same shape in the centre. But $\dhc$ takes different values for these two and as a result this fails,
especially when profiles with tails are included, as depicted in Figure \ref{fail531}. 
This also shows that investigating limited types of shapes only, as has been done in the previous works, is insufficient to reveal 
the PBH formation condition which is generally accurately applicable to various profiles.

In order to obtain reliable measures to judge if PBHs are formed or not,
we must introduce parameters which can separate the effects of the
central part collapsing into holes and those of the surrounding tail
part.
Let us define $r_p(0<p<1)$ by 
\be
K_\mathrm{i}(r_p)=p.
\ee 
After a number of trials of combinations
\be
\Delta_{p,q}\equiv r_p-r_q,
\ee
\be
I_{s,j}\equiv \int^{r_s}_0r^jK_\mathrm{i}(r)dr,
\ee
where $p<q$ and $j=0,1,2$,
it turned out that a relatively clear separation between configurations which collapse to PBHs and
those
which do not 
is obtained by the following combination:
\be
\Delta\equiv r_{1/6,5/6}=r_{1/6}-r_{5/6}
\ee
%\be
%K_\mathrm{i}(r_{3/5})=3/5, \quad K_\mathrm{i}(r_{5/6})=5/6, \quad K_\mathrm{i}(r_{1/6})=1/6, 
%\ee
and 
\be
I\equiv I_{3/5,2}=\int^{r_{3/5}}_0r^2K_\mathrm{i}(r)dr.
\ee
The quantity $I$ is similar to $\dhc$ in the sense that it is a
volume integral of curvature profile closely related to the density 
fluctuations (see eq.\ (\ref{deltabarfirst})), but the important difference is
that it is not affected by the tail part of the overdense region, since
the upper limit of the integral is cut off at $r_{3/5}$.
On the other hand, $\Delta$ represents the width of the transition region 
between $K_\mathrm{i}\sim 1$ and $K_\mathrm{i}\sim 0$.

\begin{figure}[h]
\begin{center}
\includegraphics[width=16cm,keepaspectratio,clip]{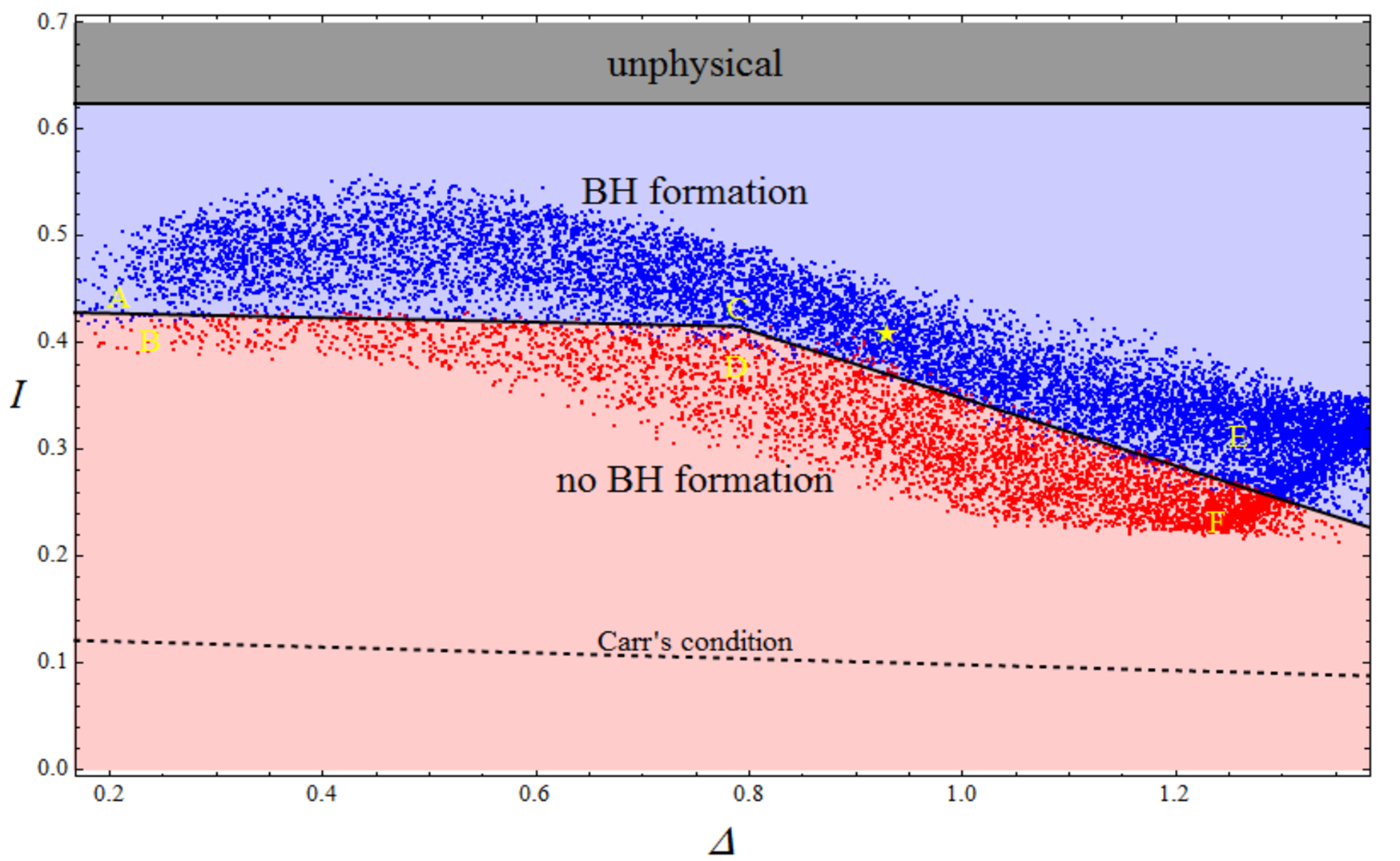}%jpgの方がmassを追加した
\end{center}
\caption{
The PBH formation condition for the profiles represented by (\ref{newfunction}). 
The blue and red points correspond respectively to the profiles which lead to the black hole formation and 
those which do not. The shaded region labelled "unphysical" corresponds to the profiles 
which do not satisfy $K_\m{i}(r)<1/r^2$ and therefore are unphysical. 
The profile used as an example of the PBH-forming cases in this paper corresponds to the yellow star in this figure.
The dashed line corresponds to the Carr's condition.
}
\label{conditionink}
\end{figure}

Figure \ref{conditionink} shows the results of numerical calculation
with various initial conditions of the five-parameter family
(\ref{newfunction}). Specifically we have chosen the values of model
parameters in (\ref{newfunction}) in the range $0 \leq A \leq 1$, 
$0 \leq B \leq 1$, $\sigma_\m{min} \leq \sigma_1 \leq \sigma_2 \leq \min\{2\sigma_1,1.14/\sqrt{1-A}\}$,
where $\sigma_\m{min}$ is chosen to search only the profiles relevant to revealing the PBH formation condition 
and
$n=1,2,3,4,5$. 
The reason for the upper bound of $\sigma_2$ is because if $\sigma_2$ is too large compared to $\sigma_1$,
the second term of the function (\ref{newfunction}) dominates and therefore this function
reduces to a single gaussian.
In each numerical simulation, the values of the five parameters are chosen randomly from the range of each parameter. 
The initial conditions are set up by the second order asymptotic expansion. 
The eventual numerical errors, estimated by the constraint equation (\ref{constraintbar}) 
have been confirmed to be less than a few ％.

\begin{figure}[h]
\begin{center}
\includegraphics[width=16cm,keepaspectratio,clip]{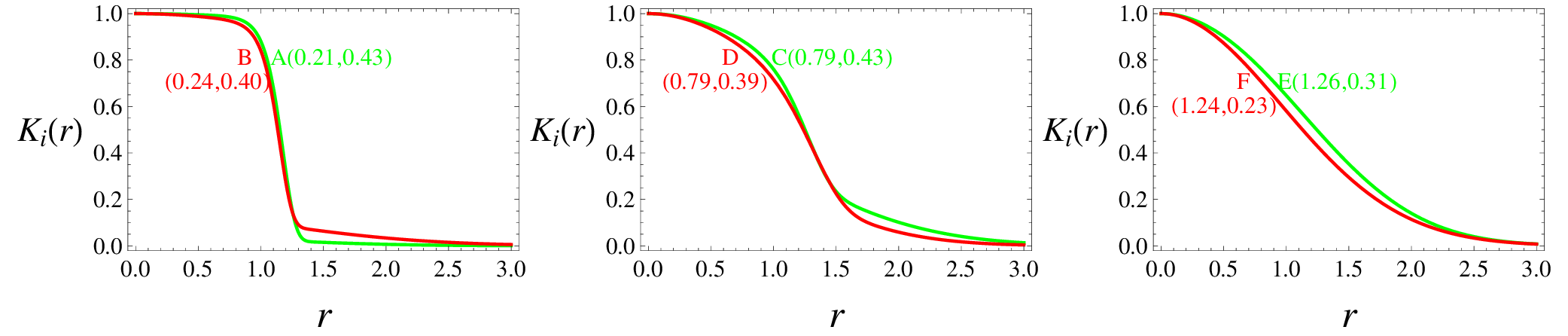}%jpgの方がmassを追加した
\end{center}
\caption{
Profiles corresponding to the points marked as A-F in Figure \ref{conditionink} with values of 
$(\Delta,I)$ for each profile.
}
\label{AtoF}
\end{figure}

As is seen there the
condition for PBH formation can be quite well described by the following fitting formula:
\be
%(-0.02 \Delta+0.43) \Theta (-(\Delta-0.79))+(-0.31\Delta+0.66)
%   \Theta (\Delta -0.79)<I
(S_1(\Delta-\Delta_{\mathrm{b}})+I_{\mathrm{b}})\Theta(-(\Delta-\Delta_{\mathrm{b}}))
+(S_2(\Delta-\Delta_{\mathrm{b}})+I_{\mathrm{b}})\Theta(\Delta-\Delta_{\mathrm{b}})<I
%0.042\Delta^3-0.32\Delta^2+0.25\Delta+0.38<I
,\label{fitting}
\ee
%%%%%%%%%%%%%%%%%%%%%%%%%%%%%%%%%%%%%%%%%%%%%%%%%%%%%%%%%%%%%%%%%%%%%%%%%5
where $\Theta$ denotes the unit step function and 
$(S_1,S_2,\Delta_{\rm{b}},I_{\rm{b}})=(-0.021,-0.32,0.79,0.41)$, which represent the slopes of 
the two lines and the position of the break. 
This formula corresponds to the lower solid line in Figure \ref{conditionink}. 
Figure \ref{AtoF} depicts the initial profiles corresponding to the
points  A through F marked in Figure \ref{conditionink} to provide
insights into the shapes of the profiles in each domain of the Figure \ref{conditionink}.

Note that for the larger values of $\Delta$, the threshold value for PBH formation $I$ is smaller.
This is  because when $\Delta$ is larger, the pressure gradients are smaller and in addition
gravity is relatively stronger even away from the centre, in which case gravity near the centre, 
measured by $I$, needs not be so large compared to cases with a smaller $\Delta$. 
Put differently, for $I \lesssim 0.43 \equiv I_\m{cr}$, profiles with a smaller
$\Delta$ do not result in PBH formation because the pressure gradient is
so large that the gravitational collapse is hindered.

On the other hand, for $I \gtrsim I_\m{cr}$ the criterion is not so sensitive to
$\Delta$ for the following reason.  Since we are using a
normalization condition $K_\m{i}(0)=1$, the configurations with a large
 $I$ must also have a large  $r_{3/5}$.  Thus those with $I \gtrsim
 I_\m{cr}$ have such a large  $r_{3/5}$ that the difference in
$\Delta$ does not affect the formation criterion much. 
 Note that $(\Delta,I)$ can distinguish between the PBH-forming profiles and
the non-forming cases much more decisively than $(\Omega,\dhc)$.
This is partly because $(\Delta,I)$ does not reflect the information near
the outer boundary around $r=r_\m{i}$ so much as $\dhc$. 

Some of the readers may be worried about the existence of some red points corresponding to
\textit{non}-PBH forming profiles in the BH formation region (the blue region), namely, false positives, and 
some blue points in the red region (false negatives). 
Note that it is in principle impossible to distinguish between two possibilities with 100％ accuracy
using only two quantities, while considering the profiles controlled by as many as five parameters.
However, we argue that this choice of quantities, $(\Delta,I)$, separates two regions fairly well with high accuracy. 
The false positives and the false negatives constitute only around 2.1％ of all 
the points in Figure \ref{conditionink}. %, which are randomly chosen relatively near the threshold line. 
That is, eq.(\ref{fitting}) 
can tell whether a PBH is formed or not with around 2.1％ accuracy. 
Note that this percentage is calculated only from the profiles appearing 
in Figure \ref{conditionink}, which are randomly generated relatively near the threshold line
to reveal PBH formation condition.
The fitting formula was obtained by minimizing this percentage. 
On the other hand, 
the points in the “mixed region” in Figure \ref{fail531} constitute around 10％ of all the points.
Therefore, the accuracy would be around 10％ for the case of $(\Omega,\bar{\delta}_{\rm{hc}})$, if
the threshold line was drawn somehow in Figure \ref{fail531}, as is done in Figure \ref{conditionink}. 
Note that Figures \ref{conditionink} and \ref{fail531} are depicted with exactly the same set of 
5-parameter choices. 

The dashed line in Figure \ref{conditionink} corresponds to the Carr's condition eq.(\ref{classical}). 
Note that the value '1/3' was obtained in the uniform hubble slicing and this value corresponds to 
$\bar{\delta}_{\rm{hc}}\simeq 0.22$ in the 
comoving slicing \cite{Harada:2013epa}.
It turned out that the profiles for which $\bar{\delta}_{\rm{hc}}\simeq 0.22$ is distributed around the dashed line in Figure \ref{conditionink}, 
so the original Carr's condition roughly corresponds to $0.1\lesssim I$.

The horizontal line in Figure \ref{conditionink}, originating from $K_{\rm{i}}(r)<1/r^2$, is obtained as follows. 
First, note that the ideal profile which gives the maximum value of $I$ for some 
$\Delta$ is $K_{\rm{i}}(r)=1(0<r<1),1/r^2(1<r<\sqrt{6/5}+\Delta),0(\sqrt{6/5}+\Delta<r)$, when 
$\Delta\leq \sqrt{6}-\sqrt{6/5}\simeq 1.35$.
In addition, when $\sqrt{5/3}-\sqrt{6/5}\simeq 0.2\leq \Delta$,
the value of $I$ for this ideal profile, namely, the maximum value of $I$ for corresponding $\Delta$, is 
$1/3+\sqrt{5/3}-1\simeq 0.62$, hence the horizontal line. 

Let us summarize this section emphasizing the key points of what has been done.
What is important and new is that 
we have investigated various shapes of profiles altogether, 
including those profiles which were not investigated in the previous work 
as well as 
those investigated previously. We formed "master variables" which decide if 
PBHs are formed or not. This means that even though our profiles now include 5 parameters, 
only two master variables are important in order to determine if a PBH forms. Such a detailed analysis has never been done before. 
Indeed note that in the previous publications the variety of shapes was restricted, 
where the profiles were usually parameterized by at most two parameters.
Figure \ref{conditionink} tells that the PBH formation condition can be simply 
described by using only two quantities 
for the whole variety of profiles, represented by 5 parameters. 
To reach our conclusion depicted in Figure \ref{conditionink}, 
we made a lot of trials and errors to find the two master variables
which we can use to clearly separate the domains of PBH formation and non-formation. 
So our result in Figure \ref{conditionink} is not merely
a re-parameterization of the previously known results but goes far beyond
them.  We stress once again that it is important to use variables which
can separate the effects of the central part and the boundary region.

%%%%%%%%%%%%%%%%%%%%%%%%%%%%%%5
%%%%%%%%%%%%%%%%%%%%%%%%%
%%%%%%%%%%%%%%%%%%%%%%%%%
%%%%%%%%%%%%%%%%%%%%%%%%%%%%%%%%%%%%%%%

In this section the time evolution of the perturbed region 
is calculated in the cosmic time slicing, 
but the problem with this slicing is that computation stops because a singularity shows up at the centre 
soon after the apparent horizon is formed. 
Thus, although this slicing serves well to determine the criteria of PBH formation, 
we cannot use it to investigate the final mass, which is essential in order to determine the mass spectrum. 
In the next section, we discuss the determination of the mass with a singularity avoided using 
what is called the null slicing.

\clearpage
%%%%%%%%%%%%%%%%%%%%%%%%%%%%%%%%%%%%%%%%%%%%%%%%%%%%%%%%%
\section{accretion onto PBHs and null slicing}
\begin{figure}[h]
\begin{center}
\includegraphics[width=14cm,keepaspectratio,clip]{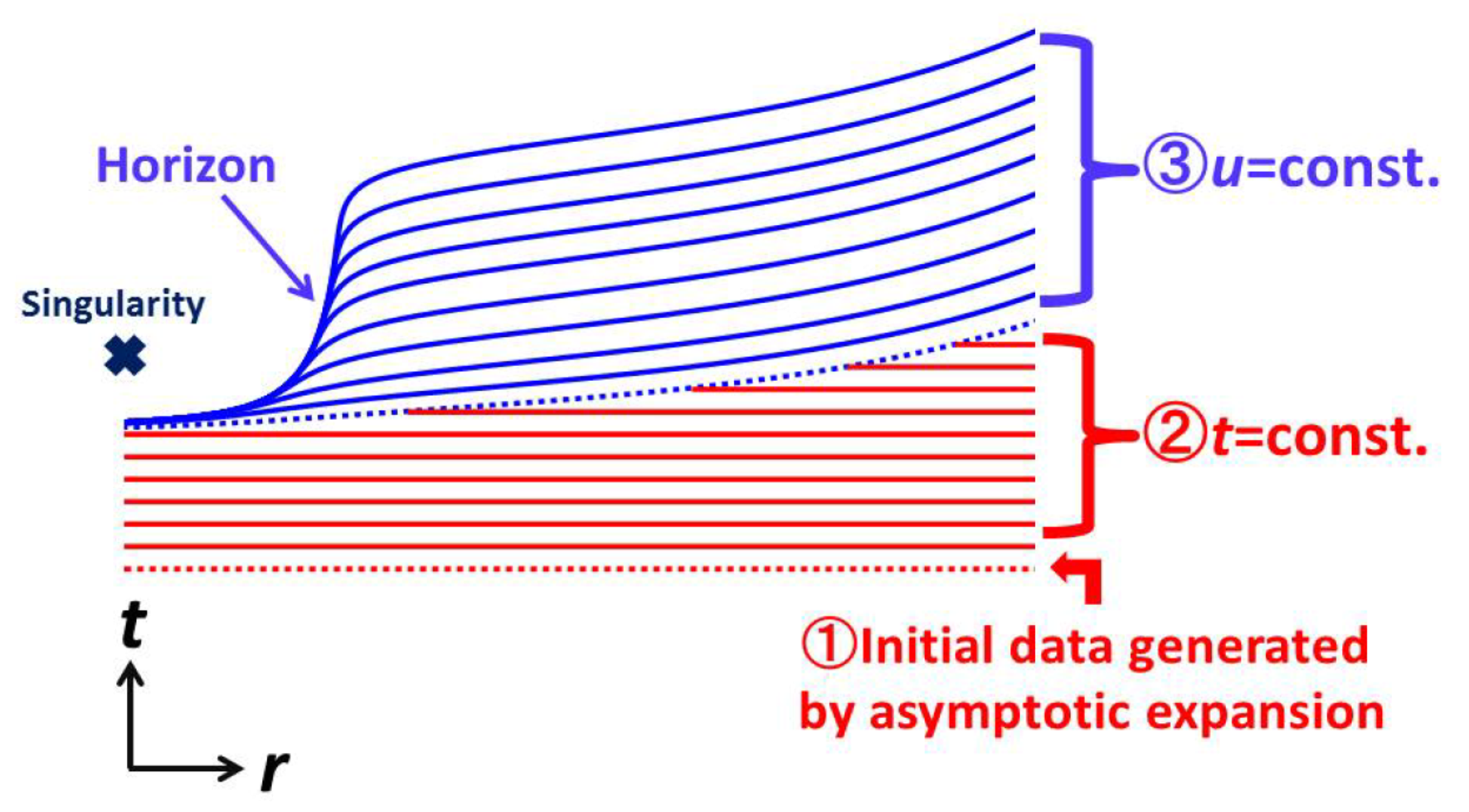}
\end{center}
\caption{Illustration of how a singularity is avoided in the null slicing. 
%One can see clearly that errors with higher-order expansions are relatively small and increase slowly 
%in comparison to those with lower-order expansions.
}
\label{slicingsink}
\end{figure}

%%%%%%%%%%%%%%%%%%%%%%%%%%%%%%%%%%%%%%%%%%%%%%%%%%%%%%%%%
The determination of the mass without facing a singularity by using 
the null slicing \cite{1966ApJ...143..452H,0264-9381-6-2-012,1995ApJ...443..717B,PhysRevD.59.124013,0264-9381-22-7-013} is discussed 
in this section. 
%so 
%numerical computation in what is called null slicing is disscused in this section, 
%which enables determination of the mass by avoiding the formation of a singularity. 
In this slicing, the space-time is sliced along the null geodesics of hypothetical photons 
emitted from the centre and reaching a distant observer. 
In other words, the space-time is sliced with the hyper-surfaces, each 
defined by a constant null coordinate $u$, the so-called observer time defined shorterly. 
By this construction of the null slicing, only the information outside the horizon is calculated, 
without looking into what happens inside the apparent horizon. 
The initial conditions are given on some hypersurface defined by $u=$const., which is depicted by a blue dotted line 
in Figure  \ref{slicingsink}, and 
are obtained using the cosmic time slicing by 
calculating the null geodesic of a hypothetical photon 
which reaches a distant observer after being emitted from the centre at some moment in time,
while at the same time recording the physical quantities of this null geodesic 
\cite{1995ApJ...443..717B}. 
As is seen in Figure  \ref{slicingsink}, 
%%%%%%%%%%%%%%%%%%%%%%%%%%%%%%%%%%%%%%%%%%%%%%%%%%%%%%%%%%%%%%%%%%%%%%%
%%%%%%%%%%%%%%%%%%%%%%%%%%%%%%%%%%%%%%%%%%%%%%%%%%%%%%%%%%%%%%%%%%%%%%
in this slicing, the information can be obtained without facing a singularity 
until a sufficiently later time 
when the eventual mass of a PBH can be determined. 

%\section{Calculation of the initial conditions in the null slicing}
%As mentioned earlier, the spacetime is sliced along the nullgeodesics 
%emanating from the center and reaching a distant observer in the null slicing. 
%Hence, the initial conditions in this slicing have to be given on some hypersurface 
%corresponding to null geodesices of photons emitter from the center at some time 
%ane reaching a distant observer. 

Let us define the time variable $u$ by first 
noting
\be
adt=bdr\label{null}
\ee
along an outgoing photon. 
Then, $u$ is defined by
\be
fdu=adt-bdr,\label{defu}
\ee
where $f$ is the lapse function and is necessary to make $du$ 
a perfect differential. 
From this definition, (\ref{null}) holds along 
the hyper-surfaces each defined by $u=$const., 
meaning that these surfaces correspond to the null geodesics of outgoing photons. 
Using $u$ as the time variable 
then means that the space-time is sliced with the null slices. 
A boundary condition on the lapse function is imposed 
by setting $a(u,r=\infty)=f(u,r=\infty)=1$, hence
$u=t$ at the surface defined by $r=\infty$. 
The physical meaning of this boundary condition is that $u$ is chosen to coincide with the proper time measured by 
a distant observer residing at spatial infinity in the background FLRW universe. 
For this reason, the null slicing is also sometimes referred to as observer time slicing in the literature. 

Let us look at the relationship between the derivatives in the 
two slicings. 
For the spatial derivatives, one finds from (\ref{defu})
\be
\fr{1}{b}\fr{\pa}{\pa r}\Big|_u=\fr{1}{b}\fr{\pa}{\pa r}\Big|_t+\fr{1}{a}\fr{\pa}{\pa t}\Big|_r.
\ee
Specifically for a function $f(t)$, depending only on $t$, 
\be
\fr{1}{b}\fr{\pa}{\pa r}\Big|_uf(t)=\fr{1}{a}\fr{\pa}{\pa t}\Big|_rf(t)\label{spderi}.
\ee
For the time derivatives one immediately finds from (\ref{null}), 
\be
\fr{\pa}{\pa u}\Big|_r=\fr{f}{a}\fr{\pa}{\pa t}\Big|_r \label{timederi}.
\ee

The Einstein equations in the null slicing were obtained in 
\cite{1966ApJ...143..452H} and 
were later used to simulate the gravitational collapse followed by 
the formation of a black hole
\cite{0264-9381-6-2-012,1995ApJ...443..717B} %bst
and recently to simulate the PBH formation as well
\cite{PhysRevD.59.124013,0264-9381-22-7-013}. 
We used numerical techniques similar to those used in 
\cite{1995ApJ...443..717B,0264-9381-22-7-013}. 
The fundamental equations are as follows:
\be
U=\fr{1}{f}R_u \label{Rdot2},
\ee
\be
\fr{1}{f}M_u=-4\pi R^2PU \label{mwrtu},
\ee
\be
E_u=-P\left(\fr{1}{\nu}\right)_u \label{Ewrtu},
\ee
\be
b=\fr{1}{4\pi \nu R^2}\label{defrho},
\ee
\be
\fr{1}{f}U_u
=-\fr{3}{2}\left(\fr{4\pi \Ga R^2}{w}P'+\fr{M+4\pi R^3P}{R^2}\right)
-\fr{1}{2}\left(4\pi \nu R^2U'+\fr{2U\Ga}{R}\right)\label{Uwrtu},
\ee
\be
\fr{1}{f}\left(\fr{1}{\nu}\right)_u
=\fr{1}{\nu\Ga}
\left(
\fr{2U\Ga}{R}
+4\pi\nu R^2U'
-\fr{1}{f}U_u
\right)\label{rhowrtu},
\ee
\be
\fr{1}{b}\left(\fr{\Gamma+U}{f}\right)'=-4\pi R\fr{\rho+P}{f}\label{fderi},
\ee
%\be
%\dot{s}\bar{U}=\fr{1}{a}(\dot{S}\bar{R}+S\dot{\bar{R}})
%\ee
where the subscript $u$ denotes differentiation with respect to $u$. 
Let us rewrite these equations in terms of the 
barred variables, defined previously. 
%as was done in the previous section.  
To do this, it is better to rewrite the $u$ derivatives by the $t$ derivatives since 
the barred variables were defined by factoring out $S(t)$ and $\rho_0(t)$, 
depending only on $t$. %$u$ derivatives have to be rewritten by $t$ derivatives 
%in order to rewrite the equations in terms of the barred variables. 
For example, using (\ref{timederi}) and the definitions of the barred variables, 
(\ref{mwrtu}) can be rewritten as
%\be
%\fr{1}{\bar{a}}\fr{\pa}{\pa t}(S^3\rho_0\bar{M})
%=-4\pi S^2\bar{R}^2\rho_0\bar{p}\dot{S}\bar{U}
%\ee
\be
(S^3\rho_0)^{\cdot}+S^3\rho_0\dot{\bar{M}}
=-4\pi S^2\dot{S}\rho_0\bar{a}\bar{R}^2\bar{P}\bar{U},
\ee
so
\be
\dot{\bar{M}}
=-4\pi \fr{\al}{t}\bar{a}\bar{R}^2\bar{P}\bar{U}
-(\fr{3\al}{t}-\fr{2}{t})\bar{M},
\ee
which then yields
%Using (\ref{ttotau}) and (\ref{S0anddefofalpha}), 
\be
\bar{M}_\tau=-4\pi \bar{a}\bar{R}^2\bar{P}\bar{U}
+3\gamma \bar{M},
\ee
with the help of (\ref{S0anddefofalpha}) and (\ref{ttotau}). 
Similarly, (\ref{Rdot2}) leads to
\be
\bar{R}_\tau=\bar{a}\bar{U}-\bar{R},
\ee
%\be
%\fr{\pa\bar{M}}{\pa \tau}=-4\pi\bar{a}\bar{R}^2\bar{P}\bar{U}+3\ga \bar{M}
%\ee
and also, noting (\ref{ttotau}) and (\ref{timederi}), (\ref{Ewrtu}) gives
\be
\bar{E}_\tau=-\bar{P}\left(\fr{1}{\bar{\nu}}\right)_\tau.
\ee
Equation (\ref{defrho}) leads to
\be
\bar{b}=\fr{1}{4\pi \bar{\nu} \bar{R}^2}.
\ee
Using (\ref{spderi}) and the definitions of the barred variables, 
(\ref{rhowrtu}) can be transformed to give
\be
\fr{1}{\bar{a}}\left(\fr{S^3}{\bar{\nu}}\right)^{\cdot}
=\fr{S^3}{\bar{\nu} \bar{\Gamma}}
\left(
\fr{2\dot{S}\bar{U}\bar{\Gamma}}{S\bar{R}}
+\fr{1}{\bar{a}}\ddot{S}\bar{U}
+\fr{\dot{S}}{S\bar{b}}\bar{U}'
-\fr{1}{\bar{a}}(\dot{S}\bar{U})^{\cdot}
\right),
\ee
which can be further rewritten using (\ref{timederi}) 
as 
\be
(\log\bar{\nu})_\tau
=3-\fr{2\bar{a}\bar{U}}{\bar{R}}
+\fr{1}{\bar{\Ga}}
\left(
-\fr{\bar{a}}{\bar{b}}\bar{U}'
+\dot{S}\fr{\pa \bar{U}}{\pa\tau}
\right).
\ee
%%%%%%%%%%%%%%%%%%%%%%
Using (\ref{S0anddefofalpha}), (\ref{frie}), (\ref{spderi}) and (\ref{timederi}), 
(\ref{Uwrtu}) eventually gives 
\begin{eqnarray}
\bar{U}_\tau
=-\bar{U}\fr{\al-1}{\al}
-\fr{3}{2}\bar{a}
\left[
\fr{4\pi\bar{\Ga} \bar{R}^2}{\bar{w}}
\left(
\fr{1}{\dot{S}^2}\bar{P}'
-\fr{2}{\al}\fr{1}{\dot{S}}\fr{\bar{b}}{\bar{a}}\bar{P}
\right)
+\fr{3}{8\pi}\fr{\bar{M}+4\pi \bar{R}^3\bar{P}}{\bar{R}^2}
\right.\nonumber
\\
\left.
+\fr{1}{3}
\left(
4\pi\bar{\nu} R^2
\left(
\fr{\bar{b}}{\bar{a}}\fr{\al-1}{\al}\bar{U}
+\fr{1}{\dot{S}}\bar{U}'
\right)
+2\fr{1}{\dot{S}}\fr{\bar{U}\bar{\Ga}}{\bar{R}}
\right)
\right].
\end{eqnarray}
%%%%%%%%%%%%%%%%%%%%%%%%%%%%%%
From the definitions of the barred variables and the Friedmann equation (\ref{frie}),
(\ref{fderi}) can be rewritten as
\be
\left(\fr{\bar{\Gamma}+\dot{S}\bar{U}}{f}\right)'
=-\fr{3}{2}\dot{S}^2\bar{b}\bar{R}\fr{\bar{\rho}+\bar{P}}{f}.
\ee

The equations in terms of the barred variables shown above 
include $\dot{S}$, which can be calculated by $\dot{S}=1/\sqrt{\ep r_\mathrm{i}^2}$, 
the relation obtained from (\ref{defofep}). 
%$\dot{S}$
Observe that $t$, the time variable in the cosmic time slicing, 
depends on both $u$ and $r$ ($t=t(u,r)$) in the null slicing, 
so do $\ep$ and $\tau$. 
Then it is necessary to determine 
$\ep$ and $\tau$ at a point $(u,r)$. % and $r$. 
To this end, note that on the surface $r=$const., we find 
\be
dt(u,r)=\fr{f(u,r)}{a(u,r)}du=\fr{f(u,r)}{a(u,r)}dt(u,r=\infty),
\ee
which can be rewritten as
\be
d\tau(u,r)=\fr{f(u,r)}{a(u,r)}
\left(
\fr{\ep(u,r=\infty)}{\ep(u,r)}
\right)^{1/\beta}
d\tau(u,r=\infty)\label{dtau}.
\ee
Here the definition of $\ep$ (\ref{defofep}) and that of $\tau$ (\ref{deftau}) 
have been used. 
The time step $d\tau(u,r=\infty)$ 
needs to be chosen to satisfy 
the CFL condition (see the Appendix), 
and the general $d\tau(u,r)$ %, the time step at a point $(u,r)$, 
can be calculated from (\ref{dtau}). 

%\be
%R=R_0\ti{R}=rS\ti{R}\equiv S\bar{R},
%\ee
%\be
%a=\ti{a}=\bar{a},
%\ee
%\be
%u=u_0\ti{u}=\dot{S}r\ti{u}\equiv \dot{S}\bar{u},
%\ee
%\be
%e=\rho_0\bar{\rho},
%\ee
%\be
%m=\fr{4}{3}\pi \rho_0R^3\ti{\mu}=\fr{4}{3}\pi \rho_0S^3\bar{R}^3\ti{\mu}\equiv \rho_0S^3\bar{M},
%\ee
%\be
%b=\fr{R'}{\sqrt{1-Kr^2}}=\fr{S\bar{R}'}{\sqrt{1-Kr^2}}\equiv S\bar{b},
%\ee
%\be
%\nu=\fr{1}{4\pi R^2b}=\fr{1}{4\pi S^3\bar{R}^2\bar{b}}\equiv S^{-3}\bar{\nu},
%\ee
%\be
%\Ga=4\pi \nu R^2R_\mu=4\pi S^{-3}\bar{\nu} S^3\bar{R}^2\bar{R}_\mu=4\pi\bar{\nu}\bar{R}^2\bar{R}_\mu\equiv \bar{\Ga},
%\ee
%\be
%P\equiv \rho_0\bar{P},
%\ee
%\be
%w=E+\fr{P}{\nu}=\fr{e}{\nu}+\fr{P}{\nu}=\fr{\rho_0\bar{\rho}}{S^{-3}\bar{\nu}}+\fr{\rho_0\bar{P}}{S^{-3}\bar{\nu}}=\rho_0S^3\left(\fr{\bar{\rho}}{\bar{\nu}}+\fr{\bar{P}}{\bar{\nu}}\right)\equiv \rho_0S^3\bar{w}.

%\ee

The boundary conditions are imposed the same as in the cosmic time slicing 
by setting $\bar{u}=\bar{R}=\bar{M}=0, \Ga=1$ at the centre 
and $\bar{a}=\bar{\rho}=1$ on the outer boundary so that 
the numerical solution coincides with the background FLRW solution there. 

%%%%%%%%%%%%%%%%%%%%%%%%%%%%%%%%%%%%%%%%%%%%%%%%%%%%%%%%%%%%%%%%%5%

\begin{figure}[h]
\begin{center}
\includegraphics[width=14cm,keepaspectratio,clip]{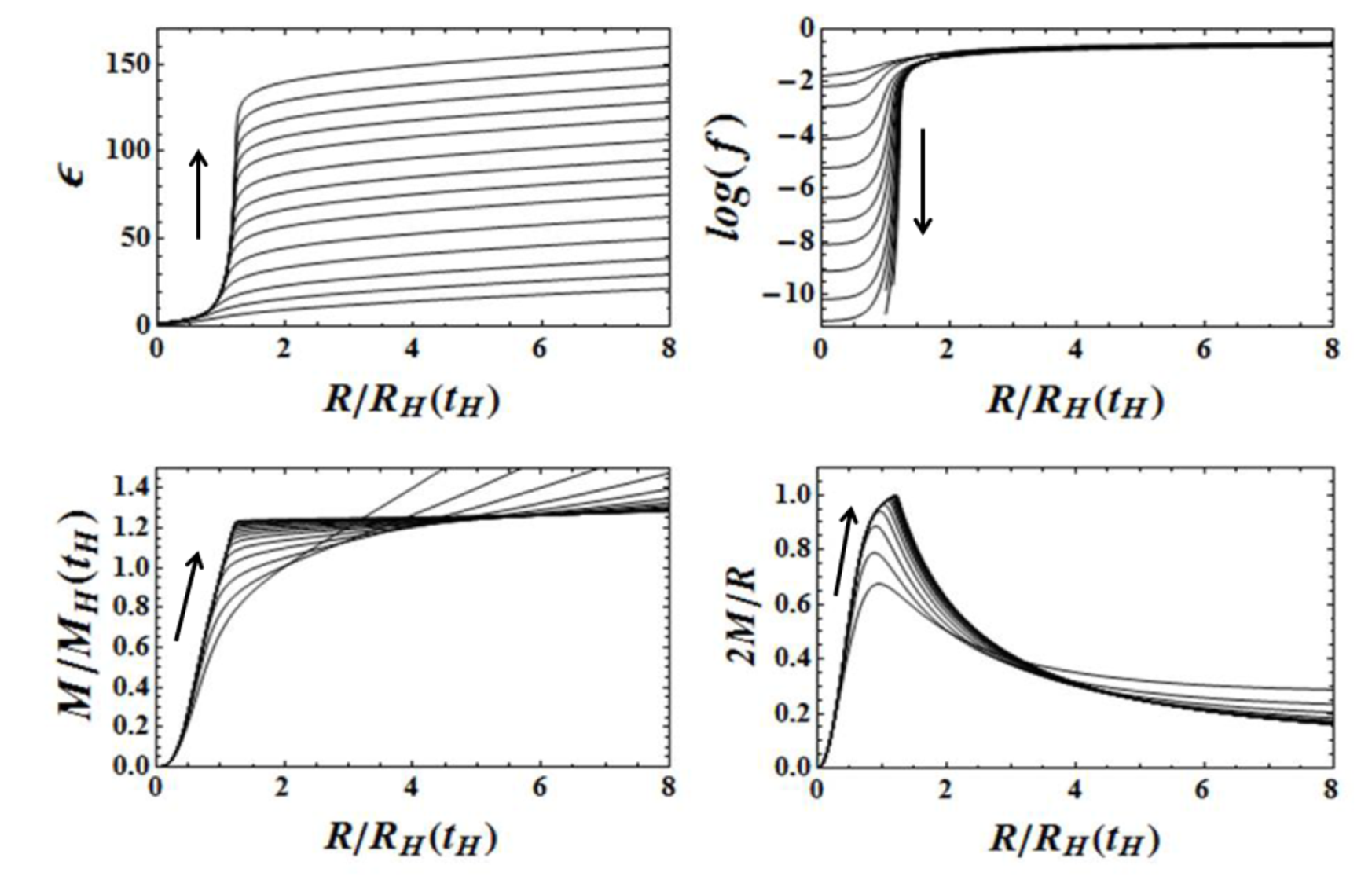}
\end{center}
\caption{
Examples of time evolutions of the relevant quantities in the null slicing for the case where $(B,\sigma)=(1,0.7)$. 
The horizontal line represents the circumferential radius $R$, normalized by the Hubble radius at the horizon crossing. 
The $u=$const. surfaces are shown in the top left panel. 
Shown in the top right panel is the lapse function, which goes to zero near the centre. 
The mass $M$ normalized by the horizon mass at the time of the horizon crossing is shown in the down left panel. 
The profile of $2M/R$, the height of which asymptotes to unity, is shown in the down right panel.
Arrows in each figure indicate the evolution in time sequence. 
}
\label{ns_figures}
\end{figure}

We now present results of numerical computations using the null slicing. 
The hypersurfaces of $u=$const., corresponding to null geodesics, are shown in the top left panel of 
Figure  \ref{ns_figures}. Observe that the intervals between the null geodesics 
are tiny in the central region, meaning that time does not pass there in effect. %is stopped their. 
Therefore, the formation of a singularity can be avoided in this slicing as expected. 
The upper lines in this figure correspond to the null geodesics 
of the hypothetical photons which are emitted from the centre at later times and 
feel the effects of stronger gravity, 
so that they 
need more time to 
reach a distant observer. 
In this figure 
there is an envelope curve of the null geodesics, which 
 shows approximately the location of the apparent horizon. 
In this way the time evolution is computed only outside the apparent horizon, 
so the eventual mass of a PBH can be determined without facing a singularity. 
From the same figure, the apparent horizon radius can be confirmed to asymptote to a constant value after its formation. 
This means that the black hole mass asymptotes to a constant value because 
$R=2M$ on the apparent horizon, 
and this behavior of the mass can be confirmed by the converging curves of the mass profile in the 
bottom left panel of Figure \ref{ns_figures}. 

The flatness of the mass profile in later time can be understood by noting 
that the energy density in a region away from the centre decreases due to the expansion of the universe 
and also due to the existence of an underdense region surrounding the central overdense region. 
As mentioned earlier, this behaviour of the mass profile is similar to that of the 
vacuum inhabited by a star at the centre, in which case the mass profile is a monotonically increasing function in $r$ 
inside the star and is flat outside.

The time evolution of $2M/R$ is shown in the bottom right panel of Figure  \ref{ns_figures}. 
The fact that the height of the peak of $2M/R$ saturates to unity in a late time in the null slicing 
is clearly seen, which is contrasted with the cosmic time slicing in which 
the peak is confirmed to exceed unity 
(see Figure  \ref{cs_figures}). 
This feature of the time evolution being frozen near the centre in the null slicing can also be confirmed by 
the behaviour of the lapse function, 
which is shown in the top right panel of Figure  \ref{ns_figures}. 
Note also that the spatial derivatives of $f$ become large near the apparent horizon, 
which shows the necessity of 
Adoptive-Mesh-Refinement(AMR), described in the Appendix. 
Thus Figure  \ref{ns_figures} as a whole shows that using the null slicing the Einstein equations can be solved until a 
sufficiently late time when the eventual mass of a PBH can be determined. 

In this paper the mass of a PBH for one particular profile was shown
just to illustrate how the mass can be obtained by null slicing.
In our future work, as was mentioned in the Introduction, we will investigate the mass evolution and
the eventual mass of PBHs for a wide variety of profiles considered in this paper.

%%%%%%%%%%%%%%%%%
%%%%%%%%%%%%%%%%%%%%%%%%%%%%%%%%%%%%%%%%%%%%%%%%%%%%%%%%%%%%%%%%%%%%%%%
%%%%%%%%%%%%%%%%%%%%%%%%%%%%%%%%%%%%%%%%%%%%%%%%%%%%%%%%%%%%%%%%%%%%%%
%\end{document}

%\section{Conclusion, discussion and results}
\section{Conclusion}
In this paper we have presented the results of 
numerical computations of the time evolution of a perturbed 
region after the horizon 
re-entry.  The initial conditions for these numerical computations 
were given using an analytical asymptotic expansion technique developed in our previous paper. 
By calculating the time evolution of various initial 
perturbations, the condition for PBH formation has been investigated. 
We have extended preceding analyses by performing many more numerical
computations of PBH formation based on the initial curvature profiles
characterized by five parameters which not only reproduce the variety of 
profiles near the centre but also incorporate the possible extended features
in the tail region (see eq.(\ref{newfunction})). 

We have shown that the criterion of PBHs formation can still 
be expressed in terms of two crucial (master) parameters 
which correspond to the averaged amplitude of over density in the central region and the width of transition region at 
outer boundary. As is shown in Figure \ref{conditionink}, this is
the case even though our profiles are characterized by as many as five parameters.
We have also provided a reliable physical interpretation of the two-parametric 
criterion.

%The first one is the density perturbation at the time of 
%horizon crossing averaged over the overdense region, 
%which is denoted by $\dhc$ and measures how strong gravity is. 
%The second one is the absolute value of the maximum value of radial derivatives of the perturbation profile, represented by 
%$\Omega$. We found the condition for PBH formation is expressed as $\bar{\de}_\m{min}(\Omega)<\dhc$ and 
%$\bar{\de}_\m{min}(\Omega)$ is an increasing function of $\Omega$ and spans $0.45\lesssim\dhc\lesssim 0.6$. 
%That is, when $\Omega$ is large $\dhc$ has to be larger to form a PBH. 
%This implies a larger value of $\Om$ makes pressure gradients gradients more effective so that 
%it is even more difficult for gravity to dominate and create a PBH. 
%This condition will be useful in computing abundance of PBHs presicely. 

Using a null slicing we have calculated the variation of PBH mass as a result of accretion. In this paper we present only one example
corresponding to $\Ki$ given in Figure \ref{ns_figures}. For this example, the eventual mass of the PBH is approximately equal to 
the horizon mass at the time of horizon crossing. The initial configuration in this example lies at the point 
marked by the yellow star in Figure \ref{conditionink}, which is fairly far from the critical line. 
% as is predicted in カー. 
However, it has been shown that the eventual mass of a PBH can be much smaller when 
$\dhc$ is extremely close to 
$\de_\m{min}$, the minimum value required to create a black hole
\cite{PhysRevD.59.124013,Musco:2008hv}, which would fall on the critical line in Figure \ref{conditionink}. 
In our future work we will explore this issue for the various types of $\Ki$ investigated in this paper. 

%We are also planning to calculate probability distribution of curvature profiles, which 
%will eventually enable to compute abundance of PBHs combined with the result of this paper. 
Note that 
PBHs are formed only from extremely high peaks of perturbation,
corresponding 
to the tail of 
the probability distribution of primordial perturbation. 
The profiles corresponding to these peaks have been calculated
\cite{Doroshkevich,Bardeen:1985tr} 
and turned out to be approximately spherically symmetric and monotonic near the peak. 
So in this paper we have considered only the spherically symmetric
profiles 
which are monotonic near
the centre.
However, deviation from sphericity to some extent is expected so
we will explore effects of non-sphericity in our future work as well.
% to set up the initial
%condition for the numerical analysis of 
%PBH formation in an optimal way with the help of the
%asymptotic expansion. 

\if
In our analysis the  curvature profile
has a characteristic scale much larger than the Hubble radius initially, 
in accordance with the inflationary cosmology 
\cite{Sato:1980yn,Guth:1980zm,Starobinsky-1980} which predicts
formation of superhorizon-scale curvature perturbations
\cite{Mukhanov:1982nu,Guth:1982ec,Hawking:1982cz,Starobinsky:1982ee}.
This includes those perturbations which could lead to PBH formation
 \cite{PhysRevD.54.6040,
PhysRevD.42.3329,PhysRevD.50.7173,yokoyama-1997-673,
PhysRevD.58.083510,Jun'ichi1998133,kawasaki-1999-59,
PTPS.136.338,1475-7516-2008-06-024,PhysRevD.59.103505,
PhysRevD.63.123503,
PhysRevD.64.021301,Kawasaki:2007zz,Kawaguchi:2007fz}.

In order to relate the mass spectrum of PBHs (constrained by existing observations)
with the parameters of inflationary models, in future papers we are 
planning to calculate probability of realization of the curvature profiles discussed above.
After that using PHBs as a tool to investigate the small-scale structure in the very 
early universe we will be able to obtain new observational constraints of inflationary models.
%In a future paper we plan to calculate the probability of
%realization of the curvature profiles discussed above. 
%Then, combined with the result of this paper and next paper devoted to accretion onto PBHs, 
%we will eventually be able to relate the mass spectrum of PBHs
%with the parameters of inflationary models.
\fi
\section*{ACKNOWLEDGMENTS}
This work was partially
supported by JSPS Grant-in-Aid for Scientific Research 23340058 (J.Y.), 
Grant-in-Aid for Scientific Research on Innovative Areas No. 21111006 (J.Y.),
Grant-in-Aid for Exploratory Research No. 23654082(T.H.),
and Grant-in-Aid for JSPS Fellow No. 25.8199 (T.N.).
TN thanks School of Physics and Astronomy, Queen Mary College, University of London
for hospitality received during this work.
We thank B. J. Carr for useful communications.
TN acknowledges H. Kodama, K. Kohri, K. Ioka and H. Takami for helpful comments.
%%%%%%%%%%%%%%%%%%
\section*{Appendix}
%%%%%%%%%%%%%%%%%%
In this Appendix, a few details of the techniques used in the numerical computation of this paper
are discussed.
\subsection{Determination of time steps satisfying Courant-Friedrichs-Lewy condition}
In order to guarantee numerical stability, 
the following Courant-Friedrichs-Lewy condition (CFL condition）has to be maintained during the entire computation:
\be
c_\m{s}<\fr{\Delta r}{\Delta t},
\ee
where $c_\m{s}$ is the sound velocity, which is $1/\sqrt{3}$ during the 
radiation-domination, and 
$\Delta r$ and $\Delta t$ represent intervals of gridpoints and timesteps respectively. 
In general relativity, this CFL condition is modified as follows:
\be
c_\m{s}<\fr{b(t,r)\Delta r}{a(t,r)\Delta t},\label{CFL}
\ee
which is rewritten in terms of the quantities introduced previously as%using $\tau$ and $\ep$ to yield
\be
\Delta\tau<\fr{2\sqrt{3}(1-\al)\bar{b}(t,r)}{\beta r_i\sqrt{\ep}\bar{a}(t,r)}\Delta r.
\ee
%The time step $d\tau(u,r=\infty)$ %\t{r=infで統一か}
%needs to be chosen to satisfy 
%the CFL condition, expressed in terms of $u$ as 
In the null slicing (\ref{CFL}) is rewritten to be
\be
c_\m{s}<\fr{b(u,r)\Delta r}{f(u,r)\Delta u+b(u,r)\Delta r}.
\ee
%which can be obtained by rewriting %the CFL condition 
%(\ref{CFL}) in the previous section. 
This equation gives a condition for the time interval
\be
\Delta u<\fr{1-c_\m{s}}{c_\m{s}}\fr{b(u,r)}{f(u,r)}\Delta r,
\ee
which can be reexpressed noting $\Delta u=\Delta t(u,r=\infty)$ and (\ref{ttotau}) as 
a condition for $\Delta\tau(u,r=\infty)$:
\be
\Delta\tau(u,r=\infty)<\fr{2(\sqrt{3}-1)(1-\al)}{\beta r_i}\fr{\bar{b}(u,r)}{f(u,r)\sqrt{\ep(u,r=\infty)}}\Delta r.
\ee
The time step $\Delta\tau(u,r=\infty)$ has to be chosen to satisfy this condition 
and general $\Delta\tau(u,r)$ %, the time step at a point $(u,r)$, 
can be calculated from (\ref{dtau}).

%%%%%%%%%%%%%%%%%%%%%%%%%%%%%%%%%%%%%%%%%%%%%%%%%%%%%%%%%%%%%%%%%%%%%%%

%%%%%%%%%%%%%%%%%%%%%%%%%%%%%%%%%%%%%%%%%%%%%%%%%%%%%%%%%%%%%%%%%%
%%%%%%%%%%%%%%%%%%%%%%%%%%%%%%%%%%%%%%%%%%%%%%%%%%%%%%%%%%%%%%%%%%%%%
\subsection{Numerical techniques employed in the calculation in the null slicing}
The following three numerical techniques are employed 
to minimize the amount of calculations while maintaining accuracy. 

The first one is to stop computing the time evolution in the central region 
where time is frozen. 
As we have seen, $f\rightarrow 0$ (meaning $d\tau\rightarrow 0$ from (\ref{dtau})) near the centre, 
in the cases where a black hole is formed. 
So time stops in the central region, in which case the time evolution 
does not need to be followed solving the differential equations there. 
%Suppose $f$ is sufficiently small in the central region at $u=u_1$. 
%Then, the time evolution ceases to be followed inside some radius $r_\mathrm{in}$, 
%which is defined as the maximum value of $r$ satisfying $f(u_1,r)<c_1$, 
%where $c_1$ is a constant which is sufficiently small (say, $10^{-6}$). 
%After $u_1$, the inner boundary conditions are given at $r=r_\mathrm{in}$ 
%using the fact that the physical quantities become independent of time $u$ 
%after $u_1$ in the central region where $r<r_\mathrm{in}$. 

%%%%%%%%%%%%%%%%%%%%%%%%%%%%%%%%%%%%%%%%%%%%%%%%%%%%%%%%%%%%%%%%%%%
The second technique is to choose the location of the 
outer boundary optimally during computation. 
Mathematically, the boundary conditions are imposed so that 
the solution coincides exactly with the background FLRW solution 
at spatial infinity, but in numerical computation the boundary is placed at some finite radius. 
The location of the outer boudary has to be 
chosen to be sufficiently far from the centre so that 
the numerical solution of the Einstein equations is sufficiently close to 
the solution of the FLRW universe 
since otherwise errors accumulate at the outer boundary and 
numerical computation breaks down. 
Suppose at some time $u_1$ 
the outer boundary is placed at $r=r_\m{b}(u_1)$, 
where  the following inequality is satisfied:
\be
|\bar{\rho}(u_1,r)-1|<c_1\label{cond},
\ee
with $c_1$ being a constant which is sufficiently small (say, $10^{-6}$). 
%Note that $r_\m{b}$ depends on $u_1$: $r_\m{b}=r_\m{b}(u_1)$. 
%Let us write $r_\m{b}=r_\m{b}(t_1)$ to 
Here, 
it is important to note that 
the central perturbed region gradually expands outward, 
as is shown in Figure  \ref{cs_figures} for example. 
As a result, at a later time $u_2(>u_1)$, $r_\m{b}(u_1)<r_\m{b}(u_2)$ so 
(\ref{cond}) would not hold at $u_2$ if we took $r_\m{b}=r_\m{b}(u_1)$. 
If on the other hand $r_\m{b}$ is fixed to be $r_\m{b}(u_\m{f})$, 
where $u_\m{f}$ is the final time of the numerical computation, 
(\ref{cond}) always holds during the computation. 
However, this choice leads to an unnecessarily large amount of calculations. 
%In this case, at $u_1(<u_\m{f})$ for instance, 
%even though the solution can be well approximated by the background FLRW solution 
%in the region $r_\m{b}(u_1)<r<r_\m{b}(u_\m{f})$, 
%the differential equations are numerically solved in this region, 
%the procedure which is unnecessary. 
This situation can be ameliorated by changing $r_\m{b}$ optimally by taking the following procedures. 
%setting $r_\m{b}$ 
%to be sufficiently large at the initial time and then 
%increase $r_\m{b}$ at certain intervals taking the following procedures. 
Suppose (\ref{cond})is satisfied at $u_1$ so 
$r_\m{b}$ is chosen to be $r_\m{b}=r_\m{b}(u_1)$ and 
later (\ref{cond}) breaks at $u_2$. 
Then, at this moment $r_\m{b}$ is redefined to be $r_\m{b}=r_\m{b}(u_2)$ 
and the numerical solution in the region $r_\m{b}(u_1)<r<r_\m{b}(u_2)$ 
is approximated by the background FLRW solution and appended to the 
solution in the region $r<r_\m{b}(u_1)$, already calculated numerically. 
After $u_2$, the differential equations are solved in the region $r<r_\m{b}(u_2)$. 
And if (\ref{cond}) breaks again at $u=u_3$, $r_\m{b}$ is redefined 
to be $r_\m{b}=r_\m{b}(u_3)$ and the solution is approximated by the background solution 
in the region $r_\m{b}(u_2)<r<r_\m{b}(u_3)$, which is then appended to the numerical solution 
in the region $r<r_\m{b}(u_2)$. This procedure is repeated if necessary, which 
guarantees (\ref{cond}) during the entire computation and as a result 
enables a computation with a minimal number of grids 
while maintaining accuracy in the region far from the centre 
.

%%%%%%%%%%%%%%%%%%%%%%%%%%%%%%%%%%%%%%%%%%%%%%%%%%%%%%%%%%%%%%
The third technique is the reduction of grid intervals in the region 
where the spatial derivatives become large, a technique known as Adoptive-Mesh-Refinement(AMR).
As the perturbation grows, 
spatial derivatives of physical quantities become large 
in the central region, which can be confirmed from the top right panel of 
Figure  \ref{ns_figures} for instance. 
In order to maintain accuracy in such a situation, the grid intervals need to be reduced but doing so globally 
would drastically increase the number of grid points. 
Therefore, it is optimal to reduce the grid intervals in the central region only where 
the spatial derivatives become large, while the grid intervals in the region away from the centre 
are kept relatively large, since the spatial derivatives are modest there. 

\bibliography{bib205}

\end{document}